\documentclass[preprint,12pt]{elsarticle}

\usepackage{graphicx}
\usepackage{epstopdf}
\usepackage{lineno,hyperref}

\usepackage{amsmath}
\usepackage{subcaption}
\usepackage{framed} 
\usepackage{doi}
\usepackage{url}

\usepackage[utf8]{inputenc}
\usepackage[T1]{fontenc}
\usepackage{multicol} 

\usepackage{color} 

\usepackage{nomencl}
\makenomenclature
\usepackage{framed}
\usepackage{multicol}
\usepackage{etoolbox}
\makenomenclature

\usepackage{etoolbox}
\renewcommand\nomgroup[1]{%
  \item[\bfseries
  \ifstrequal{#1}{S}{Symbol}{%
  \ifstrequal{#1}{G}{Greek}}%
]}

\usepackage{setspace} 

\usepackage{longtable}

\modulolinenumbers[5]
\usepackage{amssymb}

\usepackage{wasysym}
\usepackage{siunitx}
\usepackage{color}
\usepackage{soul}
\usepackage[framemethod=tikz]{mdframed}
\usepackage{lipsum}
\colorlet{shadecolor}{yellow!20}
\usepackage{xcolor}
\usepackage{tcolorbox}
\tcbuselibrary{breakable}

\usepackage{amsmath}

\journal{Int. J. Mass Heat Transf.}

\begin{document}

\begin{frontmatter}

\title{Drop impact onto a heated surface in a depressurized environment}

\author[JAXA,TUT]{Ryuta Hatakenaka \corref{cor1}}
\ead{hatakenaka.ryuta@jaxa.jp}
\author[TUT]{Yoshiyuki Tagawa \corref{cor1}}
\ead{tagawayo@cc.tuat.ac.jp}
\cortext[cor1]{Corresponding author}
\affiliation[JAXA]{
 	organization={Japan Aerospace Exploration Agency, Research and Development Directorate},
 	addressline={2-1-1 Sengen},
 	city={Tsukuba},
 	postcode={305-8505},
 	state={Ibaraki},
 	country={Japan}}
\affiliation[TUT]{
 	organization={Tokyo University of Agriculture and Technology, Institute of Global Innovation Research},
 　addressline={2-24-16 Nakamachi},
 　city={Koganei},
 　postcode={184-8588},
 　state={Tokyo},
 　country={Japan}}

\begin{abstract}
We investigated the impact of a droplet on a heated surface in a depressurized environment, with a particular focus on the unique outcome observed under these conditions: \emph{magic carpet breakup}. 
This phenomenon, first reported by Hatakenaka et al. [Int. J. Heat Mass Transf., 145, 118729(2019)], describes an explosive, widespread rebound of the drop.
A newly-developed thin-film Fe-Ni thermocouple array with $20~\mathrm{nm}$ thick layers unveiled surface temperature during the \emph{magic carpet breakup}.
This high-speed surface temperature measurement was synchronized with total internal reflection (TIR) imaging.
The bubble growth and the subsequent pressure release eventually led to an explosive rebound of the drop.
The bubble grew almost linearly with a slight acceleration, significantly different from the asymptotic growth observed for the bubble on a superheated substrate in a liquid pool.
The growth rate remained low even when the surface was superheated to $\delta T \sim 60~\mathrm{K}$, but it increased sharply afterward.
The surface temperature decreased sharply as the measuring junction became wet but did not recover immediately after the ring-shaped contact region passed.
Remarkably, the study captured liquid microdroplets forming at the receding contact line of a growing bubble via a side-view camera and TIR. 
The surface temperature remained relatively low due to the evaporation of microdroplets.
The threshold for microdroplet formation is related to the bubble growth rate.
\end{abstract}




\end{frontmatter}


\section{\label{sec:level1}Introduction}
A spray impingement on a heated wall can be observed in spray cooling, gas turbines, and internal combustion engines \cite{Kim2007_Review_spray,Naber&Reitz1988}.
In a rocket engine, cryogenic propellant and oxidizer are sprayed onto the internal walls of the combustion chamber to prevent  overheating \cite{Zhang2006}. 
The heat flux increases as the wall temperature increases, but it decreases sharply after it exceeds a threshold called the critical heat flux until it reaches a minimum value at the Leidenfrost point \cite{Leidenfrost1966, Baumeister1973, Quere2013}.
The drop in cooling efficiency can be a severe problem in many processes where very hot surfaces have to be cooled quickly, for example, in metallurgical or metal working processes such as quenching \cite{Chen1992,Pola2013} or during solar panel cooling \cite{Nizetic2016, Sargunanathan2016}.
Comprehensive reviews of spray cooling phenomena can be found in \cite{Liang&Mudawar2017_Review_spray1, Liang&Mudawar2017_Review_spray2, Breitenbach&Roisman&Tropea2018}.
Although the interaction of a spray with a hot wall has been widely investigated over the past decades, the underlying physics of the involved phenomena is still not fully understood.
The current design of spray cooling systems remains empirical, which is not always desirable because of the increased cost of additional testing. 
It is also difficult to test a system that utilizes special fluids or operates under extreme conditions.

\nomenclature[01]{$\alpha_\mathrm{l}$}{thermal diffusivity of liquid}
\nomenclature[02]{$c_\mathrm{pv}$}{specific heat of vapor}
\nomenclature[03]{$c_\mathrm{pl}$}{specific heat of liquid}
\nomenclature[04]{$e_\mathrm{w}$}{thermal effusivity of substrate}
\nomenclature[05]{$e_\mathrm{l}$}{thermal effusivity of liquid}
\nomenclature[06]{$G$}{Amplifier gain}
\nomenclature[07]{$\mathrm{Ja}$}{Jakob number}
\nomenclature[08]{$L$}{specific latent heat of vaporization}
\nomenclature[09]{$p_\mathrm{ch}$}{chamber pressure}
\nomenclature[10]{$p_\mathrm{sat}$}{saturation pressure}
\nomenclature[11]{$p_\mathrm{v}$}{vapor pressure in a bubble}
\nomenclature[12]{$\rho_\mathrm{v}$}{density of vapor}
\nomenclature[13]{$\rho_\mathrm{l}$}{density of liquid}
\nomenclature[14]{$R_\mathrm{cont}$}{contact radius}
\nomenclature[15]{$R_\mathrm{b}$}{bubble radius}
\nomenclature[16]{$S_\mathrm{Fe, Ni}$}{Relative Seebeck coefficient between Fe and Ni}
\nomenclature[17]{$S_\mathrm{Fe, Pt}$}{Relative Seebeck coefficient between Fe and Pt}
\nomenclature[18]{$S_\mathrm{Ni, Pt}$}{Relative Seebeck coefficient between Ni and Pt}
\nomenclature[19]{$t$}{time}
\nomenclature[20]{$T_\mathrm{0}$}{initial substrate temperature}
\nomenclature[21]{$T_\mathrm{l}$}{initial drop temperature}
\nomenclature[22]{$T_\mathrm{cont}$}{contact temperature between drop and substrate}
\nomenclature[23]{$T_\mathrm{sat}$}{saturation temperature}
\nomenclature[24]{$T_\mathrm{C}$}{cold (measuring) junction temperature of thermocouple}
\nomenclature[25]{$T_\mathrm{H}$}{hot (reference) junction temperature of thermocouple}
\nomenclature[26]{$T_\mathrm{L}$}{Leidenfrost temperature}
\nomenclature[27]{$\tau$}{dimensionless time}
\nomenclature[28]{$U_\mathrm{0}$}{drop impact velocity}
\nomenclature[29]{$v_\mathrm{v}$}{specific volume of vapor}
\nomenclature[30]{$v_\mathrm{l}$}{specific volume of liquid}
\nomenclature[31]{$V_\mathrm{mea}$}{Measured voltage of thermocouple (amplified)}

\begin{table*}[!t]
\begin{framed}
\begin{scriptsize}
\begin{multicols}{2}
\printnomenclature
\end{multicols}
\end{scriptsize}
\end{framed}
\end{table*}

In this context, the interaction of a single drop with a hot substrate under atmospheric pressure has been extensively investigated,  providing comprehensive reviews of this phenomenon \cite{Liang&Mudawar2017_Review_drop, Yarin_Roisman_Tropea_book_2017, Breitenbach&Roisman&Tropea2018}.
The influence of impact parameters, fluid properties, substrate material, and roughness on the Leidenfrost temperature, $T_L$, has also been studied extensively. 
Many empirical correlations have been proposed in the past \cite{Liang&Mudawar2017_Review_drop}.
\citet{Bernardin2002} proposed a model for sessile drops based on classical bubble growth theory.
It was extended to impinging drops by considering the change in fluid properties due to the rise in pressure resulting from drop impact at the liquid-solid interface \cite{Bernardin2004}.
\citet{Aursand2018} studied the interfacial stability of film boiling and concluded that the vapor film collapses depending on the balance between thermocapillary instabilities and vapor thrust stabilization. 
This has yielded a prediction of $T_L$ consistent with experimental data found in the literature for various liquids.
\citet{LEE&Harth&Lohse2020_SoftMatter} measured the contact and lift-off times quantitatively via X-ray and total internal reflection (TIR) imaging, indicating that a drop can rebound even without a complete vapor layer.

The influence of surrounding gas pressure on the drop impact on isothermal substrates has been extensively studied experimentally \cite{Xu2005, Driscol_Stevens_Nagel2010, Driscoll2011, deRuiter2012, deRuiter2015_1, deRuiter2015_2, deRuiter2015_3, Li&Thoroddsen2015, Li&Hicks&Thoroddsen2017} and numerically \cite{Mandre2009, Mani2010}, leading to understanding of the pressure influence on dimple formation at the liquid-air interface, thin liquid sheet ejection, entrainment of microbubbles, and the thickness distribution of the entrained gas layer. 
Notably, contact nucleation between a liquid and solid occurs from a surprisingly large distance of more than $200~\mathrm{nm}$ \cite{deRuiter2012, deRuiter2015_2}.
This is only a few times the mean free path of the gas at the surrounding pressure, which suggests that a rarefied gas effect may play a significant role in the collapse of the gas layer. 

Drop impact on a heated surface under variable pressure involves extremely complicated physics.
However, the influence of the surrounding gas on the air-cushion effect (i.e., the lift force due to the compression of gas between the drop and substrate) should also play an important role.
Further, the changes in fluid properties (e.g., saturation temperature and vapor density) and the conditions for the onset of boiling due to changing surrounding pressure should also be considered.
These phenomena are poorly understood since experimental data for different pressure conditions are lacking. 
The changes in $T_L$ and the contact time have been studied under elevated pressure \cite{Buchmuller2014} and reduced pressure \cite{Celestini2013, Orejon2014, vanLimbeek2018}.
\citet{Breitenbach2017PRL} proposed a theoretical model of heat transfer during drop impact in the nucleate boiling regime to estimate the lifetime of an impinging drop, and the results agreed well with the literature.
Static and dynamic $T_L$ tend to follow the given fluid's saturation temperature under both elevated and reduced pressures, although dynamic $T_L$ stops decreasing then increase at an extreme decrease in surrounding pressure \cite{vanLimbeek2017Soft}. 
No pure physics-based model for the dynamic $T_L$ under elevated or reduced pressure has been developed.

\begin{figure}
  \centering
  \includegraphics[width=1.0\linewidth]{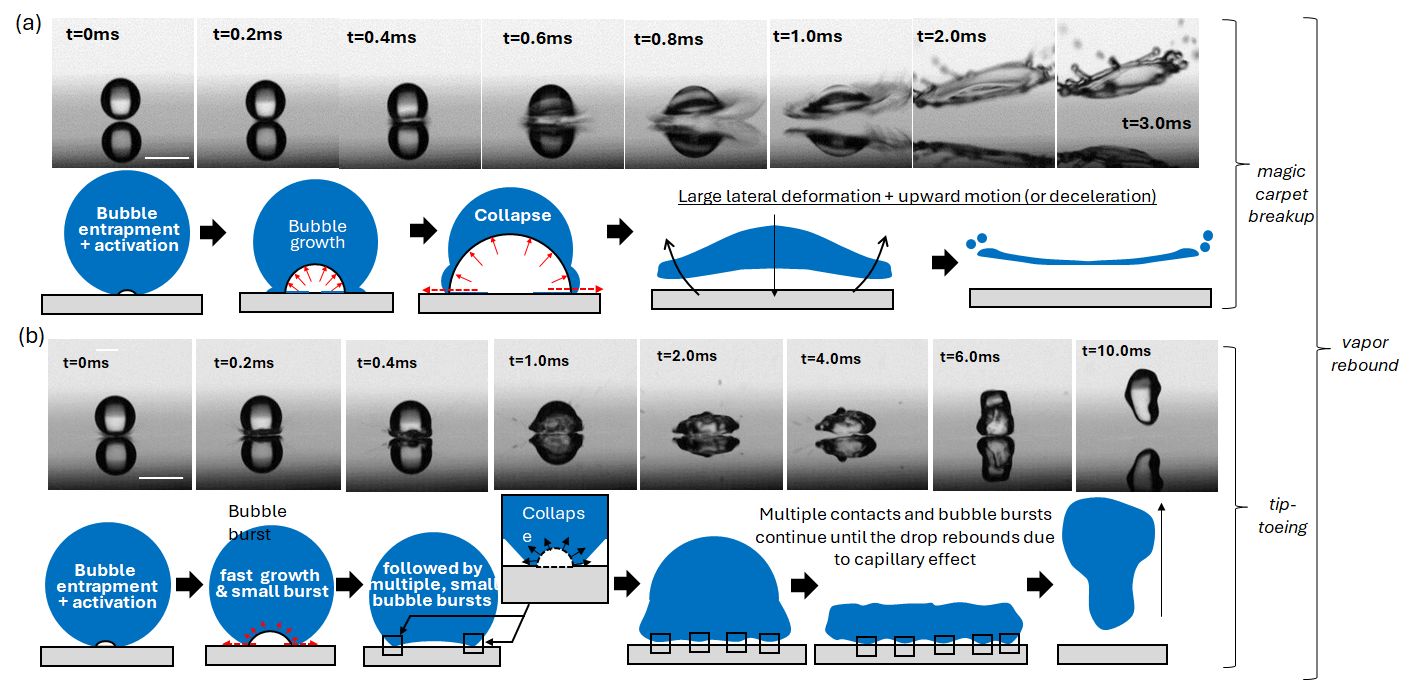}
  \caption{Experimental images and the hypothesis on the two cases of \emph{vapor rebound}: \emph{magic carpet breakup} and \emph{tiptoeing}. Pictures are taken from \citet{Hatakenaka2019}.}
  \label{Fig1}
\end{figure}

Drop impact onto a heated substrate under very low pressure (typically less than $5~\mathrm{kPa}$) leads to an explosive drop deformation and backlash motion \cite{Hatakenaka2018}.
A schematic illustration of this phenomenon is shown in Fig.~\ref{Fig1}. 
This outcome is called \emph{vapor rebound} \cite{Hatakenaka2019}, which is subdivided into two cases: \emph{magic carpet breakup}, in which the drop suddenly backlashes from the wall with a large lateral deformation, and \emph{tiptoeing}, in which attach-and-detach cycles successively occur between the drop and wall, accompanied by the generation of secondary droplets. 
\citet{Yu2019} performed TIR imaging, capturing a single bubble that grew very large. 
A theoretical model that qualitatively explains this was proposed in \cite{Hatakenaka2019}; however, no experimental data on bubble activation and growth under such conditions are in the literature. 
Further understanding of the interfacial phenomenon among the drop, substrate, surrounding gas, and generated vapor is necessary to establish a physical model to estimate drop behavior. 
\citet{Pirat2023_PRF} studied depressurization-induced breakup of a drop sitting on a heated, textured hydrophobic surface, where a bubble at the liquid-solid interface grew extensively and triggered the explosive breakup similarly to \emph{magic carpet breakup}.
They discussed the influence of the initial contact angle of the drop and proposed a theoretical model based on inertial dynamics. 
Similar explosive drop outcome have also been observed under atmospheric conditions for a water drop impacting a heated surface (e.g., titanium oxide nanotube surface \cite{Tong2017_Appl.Phys.Let.} and random silicon nanowires \cite{Auliano2018}) and for a static Leidenfrost drop made of an aqueous solution of surfactant on a smooth heated surface \cite{Moreau2019}.

Despite the importance of the thermal effect, experimental data on the temperature or heat flux during drop evaporation on a heated surface are very limited. 
The heat transfer during evaporation of a levitating drop deposited on a superheated substrate has been extensively studied (e.g., \citet{vanLimbeek2017_JFM}).
However, the heat transfer during drop impact when the drop shape changes drastically in a short time is poorly understood.
The surface temperature evolution during drop impact has been studied experimentally by several researchers \cite{Herbert&Stephan2013, Jung&Jeong&Kim2016, Qi&Weisensee2020, Schmidt&Roisman&Tropea2021} via an infrared camera, where a thin, high-emissivity layer was deposited on the IR-transparent substrate and its temperature distribution was captured from the bottom. 
This method provides the temperature data with a high special resolution, but the temporal resolution is limited (e.g., $1000~\mathrm{fps}$).
\citet{Chaze&Lemoine2017} measured the drop temperature via fluorescent imaging and pulse laser. 
This technique is unique in measuring the liquid temperature; however, the temporal resolution is limited ($10~\mathrm{Hz}$ in \cite{Chaze&Lemoine2017}).
The electrical method (e.g., resistance thermometer \cite{Seki1978} and thermocouple \cite{Pasandideh-Fard&Chandra2001,Testa&Nicotra1986}) enable very high-speed temperature measurement, although the number of measurement points tends to be very limited. 

This study investigated the impact of a droplet on a heated surface in a depressurized environment, with a particular focus on \emph{magic carpet breakup}. 
The only driving force of \emph{vapor rebound} is the difference between the bubble and the ambient pressures, where the thermodynamic balance over the entire bubble surface determines the former.
Considering previous studies on pool boiling, the heat and mass transfer around the triple-phase contact line \citep{Stephan&Hammer1994} and the liquid microlayer \citep{Carey-Textbook, Cooper&LLOYD1969, vanStralen1975_1} has significant importance. 
To understand the thermal behavior at the surface, a thin-film thermocouple array fabricated via lithography technique used in studies on pool boiling \cite{Nakabeppu2006, Tange2009, Yabuki&Nakabeppu2014} was applied to drop impact onto a heated surface.
The typical thickness of thermocouples in previous studies was in the order of several hundred nanometers, which seems sufficiently small in terms of response time; however, the bump height influences the interfacial phenomena, such as dimple formation at the bottom face, drop wetting on the surface, air-bubble entrainment, and bubble nucleation.
The thermocouple's thickness must be reduced for studies on drop impact. 
Notably, the Seebeck coefficient for bulk material found in the literature does not apply to thin-film thermocouples.
The coefficient decreases significantly when the film thickness becomes equivalent to the mean free path of the film material \citep{Salvadori2006, Liu2011_IEEE, Varrenti2011} due to surface and boundary scattering \cite{Kockert&Fischer2019, Coutts1971_Review_ThinSolidFilms}.
Thus, establishing a calibration method is critical to assure the measurement reliability of thin-film thermocouples.
It is also a challenge since a temperature gradient should be created in the wafer.
To correlate local surface temperature with drop spreading and bubble growth on the surface, high-speed TIR imaging \citep{Kolinski2012, Shirota2017, vanLimbeek2018, LEE&Harth&Lohse2020_SoftMatter} was performed simultaneously with the temperature measurement using a thin-film ($20~\mathrm{nm}$ for each electrode) thermocouple array. 
The thermocouples were calibrated by creating a thermal equilibrium with a temperature gradient between the measuring and the reference junctions.
The drop impact experiment was performed in both atmospheric and depressurized environments 

This paper describes the experimental method, including details of a newly-developed thin-film Fe-Ni thermocouple array, discussed in Sec.~\ref{sec:level2}. A calibration test of the temperature measurement system (Sec.~\ref{sec:level3.1}) and a control case study at atmospheric conditions (Sec.~\ref{sec:level3.2}) are presented. The main discussion on the experimental results in a depressurized environment is given in Sec.~\ref{sec:level4}, followed by the conclusion and remarks in Sec.\ref{sec:level5}.

\section{\label{sec:level2} Experiments: fabrication of thermocouple array and measurement system}
\subsection{\label{sec:level2.1}Fabrication of thin film thermocouple array}

\begin{figure}
  \centering
  \includegraphics[width=1.0\linewidth]{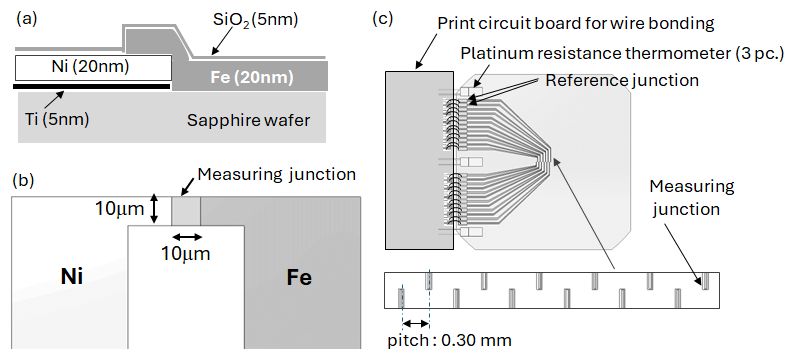}
  \caption{The layout (a) and layer constitution (b) of the Fe-Ni thin-film thermocouple array fabricated on a sapphire wafer.}
  \label{TC_array}
\end{figure}

A thin-film thermocouple array was fabricated via lithography method in Nano Processing Facility of National Institute of Advanced Industrial Science and Technology, Japan.
An overview of the array is shown in Fig.~\ref{TC_array}.
The circuit was formed on a transparent, double-side-polished sapphire wafer to enable synchronizing temperature measurement with TIR imaging.
The thickness of the wafer (500$\mu$m) was chosen so that the thermal boundary layer formed in the wafer does not reach the backside in the experimental time scale in the present study. 

The layer constitution is shown in Fig.~\ref{TC_array}(b). 
For metal alloys, the composition of the deposited layer fabricated via sputtering or vapor deposition can differ from that of the target material.
Thus, the electrodes of thermocouples should consist only of pure metal to ensure repeatability. 
This study selected a combination of iron (Fe) and nickel (Ni) based on the estimated sensitivity from the Seebeck coefficient relative to silver (Ag) in the literature.
An empirical thickness of $20~\mathrm{nm}$ was chosen for Fe and Ni layers based on trial fabrications and evaluations of different layer thicknesses.
A thin titanium layer was deposited beneath the Ni layer to improve adhesion strength. 
A $300~\mathrm{nm}$ aluminum layer was deposited on only the reference junctions for wire bonding. 
The entire wafer surface (except for the bonding pads) was coated with silicon dioxide layer for electrical insulation and anti-oxidization.
The size of the measuring junction was set to $10$ square micrometer.

A circular two-inch wafer was cut into a rectangle with rounded corners to provide straight edges.
Wire bonding pads located along a straight edge were wire-bonded to a neighboring print circuit board. 
Paired bonding pads on the wafer corresponded to the reference junction of each thermocouple channel. 
The temperature of the reference junctions was evaluated with three platinum-resistance thermometers bonded to the wafer with heat-conducting adhesive.

\subsection{\label{sec:level2.2} Amplifier for thermocouple array}
\begin{figure}
  \centering
  \includegraphics[width=1.0\linewidth]{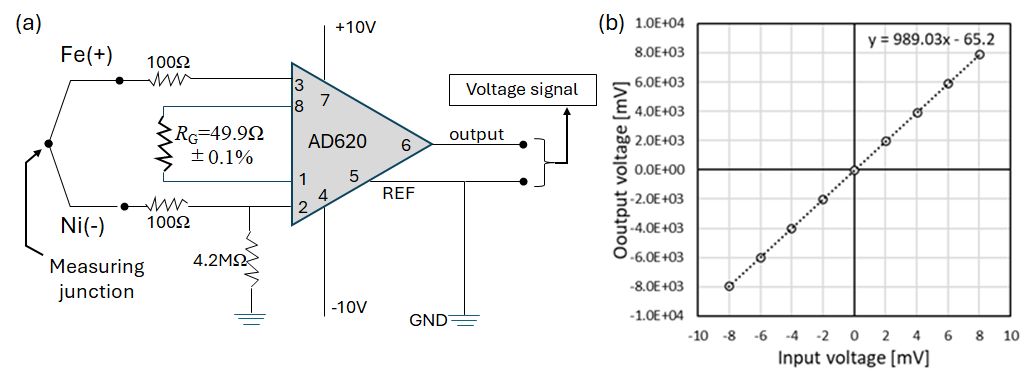}
  \caption{Amplifier for thermocouple array. (a) The circuit diagram, (b) example of calibration result.}
  \label{Amplifier}
\end{figure}

The thermocouple voltages were amplified using a custom amplifier circuit (based on AD620, Analog Devices Inc.) with a gain of 991, as shown in Fig.~\ref{Amplifier}(a).
The amplifier circuit is similar to those presented in \citep{Moghaddam2003, Tange2009}.
The gain was determined with the resistance $R_\mathrm{G}$.
The power was supplied from a DC voltage source (Kikusui, PMM35-1.2DU), whose ground pin is used as the ground shown in Fig.~\ref{Amplifier}(a). 
The amplifier gain was confirmed by applying a known, fixed voltage using a voltage source (Advantest, R6243) and showed a linear relationship between input and output voltages (Fig.~\ref{Amplifier}(b)). 
Fitting lines were drawn for twelve circuits based on the least-square method and provided an amplifier gain (i.e., the slope of the fitting line) of $990.5 \pm 1.5$. 
The variation was slightly larger than would be expected from the variation of $R_\mathrm{G}$ ($\pm 0.1\%$ corresponding to a gain of $\pm0.99$) and is attributed to variation in the performance of the amplifier IC and measurement error.
Calibration of the amplifier gain was not performed for all circuits; instead, the entire system, including TC arrays and amplifiers, was characterized in an "end-to-end" manner in the calibration test (Sec.~\ref{sec:level3.1}).     

\subsection{\label{sec:level2.3} Experimental setup}

\begin{figure}
  \centering
  \includegraphics[width=0.9\linewidth]{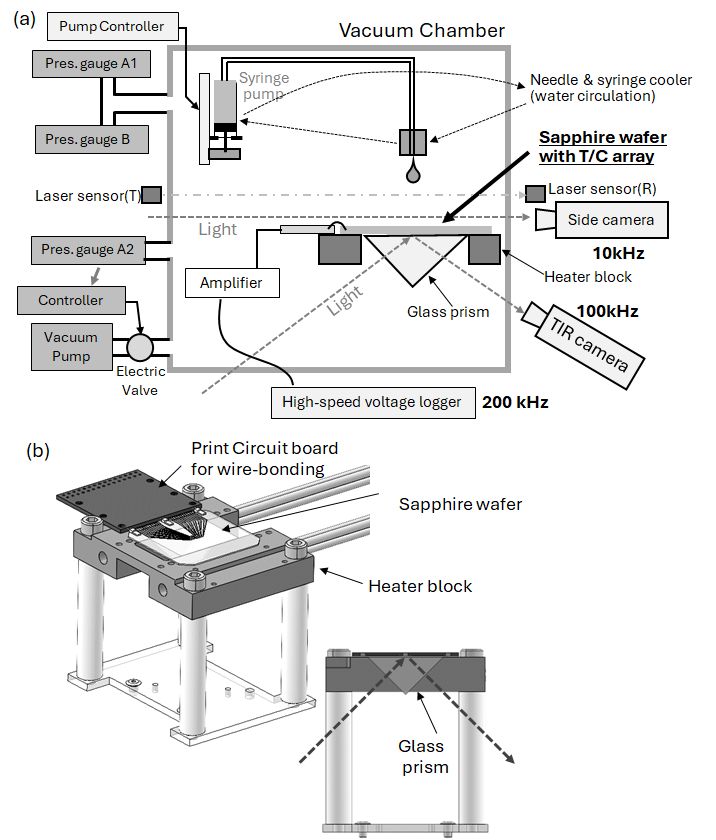}
  \caption{The experimental setup. The sapphire wafer, on which thermocouple array is fabricated, is attached to a heater block. A glass prism is attached to the bottom of the wafer with immersion oil between them. The chamber pressure is maintained at the target value by controlling the electric valve based on the measured pressure via pressure gauge A2. The pressure is measured via pressure gauges A1 (full range) and B (for low pressure) and recorded with the logger. The laser sensor generates a trigger signal commonly used in the two cameras and high-speed voltage logger.}
  \label{Setup}
\end{figure}

The experimental setup is shown in Fig.~\ref{Setup}. 
The sapphire wafer, on which the thermocouple array was fabricated, was bonded to a heater block with heat-conducting adhesive.
The block temperature was controlled with a controller and two heater rods.  
The block has a large cutout at the center, and a glass prism for TIR imaging was attached to the bottom side of the wafer with immersion oil between them.
Two high-speed cameras were used: FASTCAM SA-X2 (Photron Ltd.) for TIR and FASTCAM Mini UX50 (Photron Ltd.) for the side view. 
LEDs (HAYASHI-REPIC CO., LTD., Luminar Ace LA-HDF158A) provided illumination. 
The amplifiers for the thermocouples were located inside the chamber to minimize the noise on the raw thermocouple signals.
The amplified signals were recorded by a commercial high-speed voltage logger (Keyence, NR-500 and NR-HA08) at a rate of 200kHz. 
The chamber pressure was controlled with an oil rotary pump, a pressure gauge A2 (KYOWA ELECTRONIC INSTRUMENTS Co.Ltd., PAB-A-200KB), an electric valve, and a controller.
The pressure was also measured via separate pressure gauges A1 (PAB-A-200KB, full range) and B (MKS instrument, Inc., Balatron capacitance manometer 627F21TBC2B for less than $2.666~\mathrm{kPa}$). 
The water drop was generated by a needle and a syringe pump in the chamber.
To mitigate bubbling of water under depressurization and maintain the initial drop temperature, the syringe and the needle were enclosed by water jackets cooled via external water circulation.
The cooling water temperature was set at $10\,^\circ\mathrm{C}$.
Typical diameter of the drop was $2-3~\mathrm{mm}$. 
A laser sensor (IB-05, IB-1000, Keyence Co.) detected the drop falling from the needle, generating a trigger signal for the two cameras and the logger. 
The spatial and temporal resolutions for TIR are 18.9 $\mu \mathrm{m/pixel}$ and 16.7 $\mu \mathrm{s/frame}$ (exposure: 6.25$~\mu$s).
The time $t=0$ is defined as the moment when either a wet region (thick gray color) or a faint ring was observed.
The ring indicates the formation of a dimple on the bottom of a drop with a minimum thickness of the vapor layer within the range of the evanescent wave.

\subsection{\label{sec:level2.5} Image processing}
\begin{figure}
  \centering
  \includegraphics[width=0.9\linewidth]{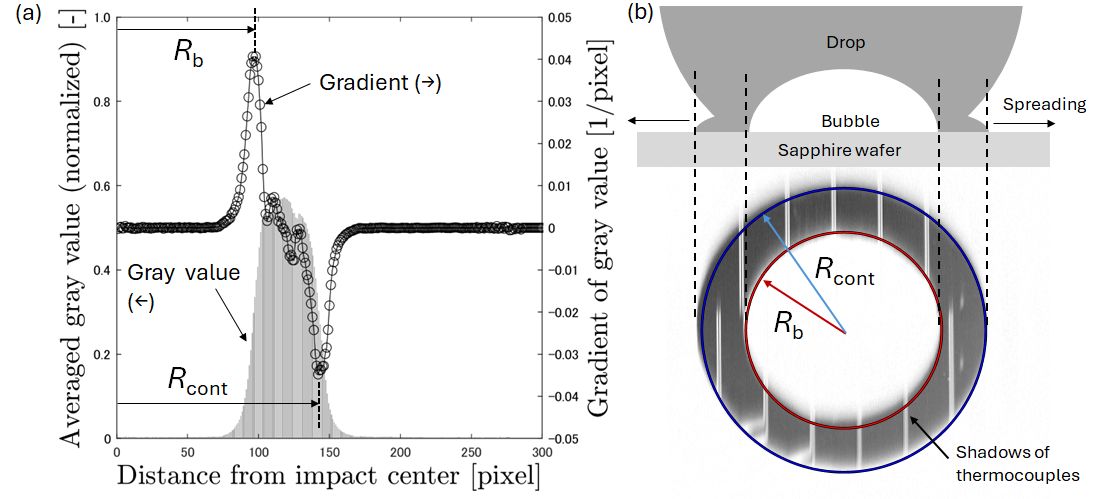}
  \caption{Image processing method. (a) Determining the contact radius $R_\mathrm{cont}$ and bubble radius $R_\mathrm{b}$ from the distribution of area-averaged gray values as a function of the distance from the impact center. (b) Exemplary fit circles with the radii of $R_\mathrm{cont}$ and $R_\mathrm{b}$ overlaid on the TIR image after background subtraction and transformation.}
  \label{ImageProcessing}
\end{figure}

The experimental images were analyzed using the MATLAB\textregistered$~$image processing toolbox.
The bubble radius $R_\mathrm{b}$ and the contact radius $R_\mathrm{cont}$ were determined using the following procedure. 
First, the coordinates of the impact center were determined in each TIR image as the center of mass of the contact region (black region) in the binarized image.
The averaged value of the coordinates in several successive images was used.
The distance $d$ from the center was calculated pixel-wise, with all pixels categorized into groups according to $d$, with an accuracy of $1~\mathrm{pixel}$. 
The pixels in a group are in a ring-shaped region.
The average gray value in each group was calculated and is shown as a function of $d$ in the bar chart in Fig.~\ref{ImageProcessing}(a).
Finally, the bar chart is smoothed by a moving average, and the gradient is calculated. 
$R_\mathrm{b}$ and $R_\mathrm{cont}$ were determined as the distance at the gray values' maximum and minimum gradient, respectively.

\section{\label{sec:level3} Calibration and control case study at atmospheric pressure}
\subsection{\label{sec:level3.1} Calibration of temperature measurement system}
\begin{figure}
  \centering
  \includegraphics[width=1.0\linewidth]{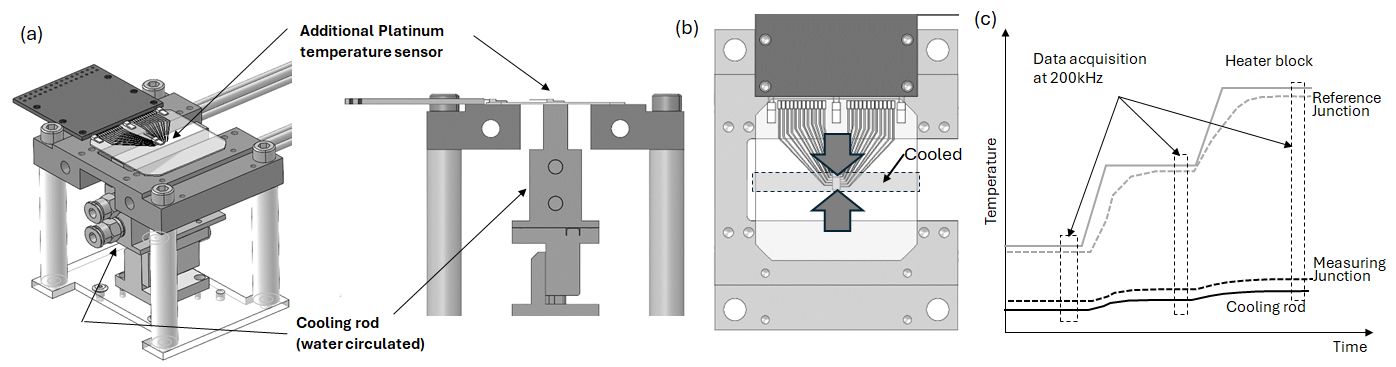}
  \caption{Configuration for calibrating the temperature measurement system. (a) Locations of the additional platinum sensor and a cooling rod, (b) one-dimensional heat flow in the sapphire wafer, (c) exemplary temperature profile.}
  \label{Calibration_Setup}
\end{figure}

The configuration of the calibration test for the temperature measurement system is shown in Fig.~\ref{Calibration_Setup}(a). 
An additional platinum thermometer was taped to the center of the sapphire wafer (at the location of measuring junctions), which was cooled continuously via an aluminum rod chilled by water circulation. 
The tip of the rod has a rectangular shape so that the entire width of the wafer is uniformly cooled. 
This configuration allows us to assume a one-dimensional heat flow from the edges of the wafer bonded to the heater block to the central region, as shown in Fig.~\ref{Calibration_Setup}(b).
The cooling water temperature was set at $10\,^\circ\mathrm{C}$, and the temperature of the heater block was changed stepwise as illustrated in Fig.~\ref{Calibration_Setup}(C).
The output signals were recorded for $0.5~\mathrm{s}$ at $200~\mathrm{kHz}$ to evaluate the high-frequency noise.

\begin{figure}
  \centering
  \includegraphics[width=1.0\linewidth]{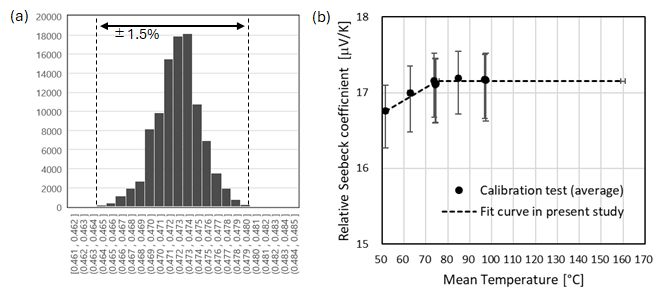}
  \caption{Results of thermocouple calibration. (a) The histogram of the output voltage after amplification for $0.5~\mathrm{s}$, during which the temperatures at both the measuring and reference junctions are constant, (b)the relationship between the relative Seebeck coefficient of the thermocouple and the mean temperature of two junctions.}
  \label{Calibration_Result}
\end{figure}

An example of data dispersion in the voltage signal after amplification is shown in Fig.~\ref{Calibration_Result}(a).
It was confirmed that random noise was suppressed within $\pm 1.5\%$. 
The measured voltage $V_\mathrm{mea}$ is determined by  $V_\mathrm{mea} = G(T_\mathrm{H}-T_\mathrm{C})S_\mathrm{Fe, Ni} = G(T_\mathrm{H}-T_\mathrm{C}) (S_\mathrm{Fe, Pt} - S_\mathrm{Ni, Pt})$, where $G$ is amplifier gain. $T_\mathrm{H}$ and $T_\mathrm{C}$ are the temperature of the reference and measurement junctions, respectively.
$T_\mathrm{H}$ was determined as the average temperature of the three platinum thermometers.
$S_\mathrm{Fe, Ni}$, $S_\mathrm{Fe, Pt}$, and $S_\mathrm{Ni, Pt}$ indicate the relative Seebeck coefficient between the two materials written in the suffix.
The bulk value of $S_\mathrm{Fe, Ni}$ was $33.5~\mu \mathrm{V/K}$, taken from the literature values of $S_\mathrm{Fe, Pt} = +18.8~\mu \mathrm{V/K}$ and  $S_\mathrm{Ni, Pt} = -14.7~\mu \mathrm{V/K}$. 

The relationship between $S_\mathrm{Fe, Ni}$ of the thin-film thermocouple and the mean temperature of the two junctions (average of $T_\mathrm{H}$ and $T_\mathrm{C}$) is shown in Fig.~\ref{Calibration_Result}(b).
In this graph, the average value of $S_\mathrm{Fe, Ni}$ among all channels is used as a representative value, while the highest and lowest values are indicated with an error bar. 
For simplification, $G=991.0$ is used for all channels.
The variation includes the measurement error of $T_\mathrm{H}$ and $T_\mathrm{C}$ and the variation in the amplifier gain, and thus does not directly indicate the actual dispersion in $S_\mathrm{Fe, Ni}$ among channels.
As expected from the literature, $S_\mathrm{Fe, Ni}$ determined in this study ($16.8-17.2 ~\mu \mathrm{V/K}$) is significantly lower than the bulk value of $33.5~\mu \mathrm{V/K}$. 
A small temperature dependency was observed. 
Although the highest temperature of the heater block in the calibration test was the same as in the drop experiment, the maximum mean temperature was $98\,^\circ\mathrm{C}$ due to the low water temperature and good thermal contact between the rod and the substrate. 
The average $S_\mathrm{Fe, Ni}$ among the relatively high-temperature region (at the mean temperature from $70$ to $100\,^\circ\mathrm{C}$) calculated for each channel was used for all temperature ranges in Sec.~\ref{sec:level3.2} and Sec.~\ref{sec:level4}.

\subsection{\label{sec:level3.2} Control case study at atmospheric pressure}
\begin{figure}
  \centering
  \includegraphics[width=1.0\linewidth]{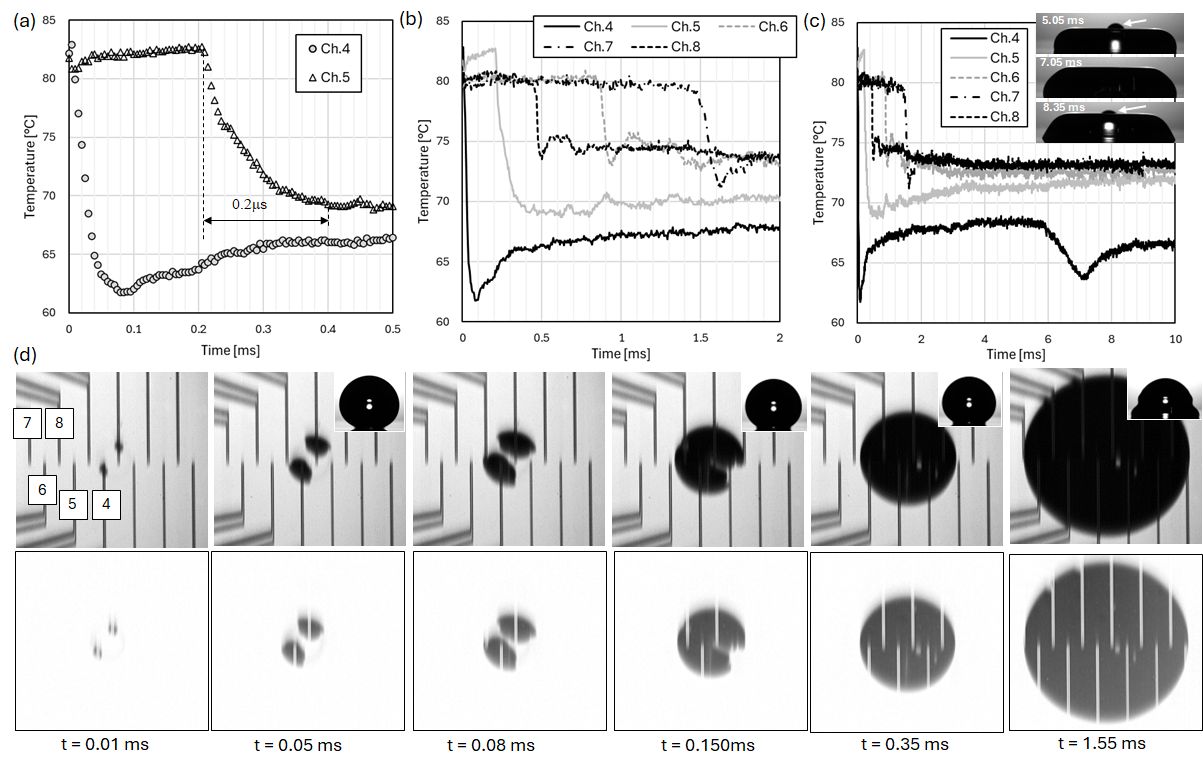}
  \caption{Temperature evolution during initial wetting and spreading of the impacting water drop. $p_\mathrm{ch} = 101~ \mathrm{kPa}$, $T_\mathrm{0}=80.9\,^\circ\mathrm{C}$ (supplemental video 1a and 1b). (a) A close-up view of the temperature trend of thermocouple Ch.~4 and Ch.~5 during initial wetting. (b) Time evolution of surface temperature at five locations during drop spreading. (c) Long-term trend of the surface temperature at the five locations. The insets are the side-view images when the second bump of Ch.~2 is observed, where the top disappears and reappears within the drop. (d) The TIR images during the wetting process ($t <= 0.150~\mathrm{ms}$) and the spreading process (($t > 0.150~\mathrm{ms}$) before background subtraction (top) and after subtraction (bottom). The insets in the top row are the side-view images at the time. The time $t~=~0$ is when the black region first appears in the TIR image. The locations of the thermocouple channels are indicated in the first photo in (d). Note that the TIR image after subtraction always becomes white at the thermocouple wires, regardless the wetting state of the location.}
  \label{ControlCase}
\end{figure}

The temperature trends and TIR image sequences during the initial stages of drop spreading are shown in Fig.~\ref{ControlCase}. Direct contact occurs over part of the faint ring after a certain delay \citep{Kolinski2012}. 
The temperature of Ch.4 drops sharply with the time scale of $10-20~\mu \mathrm{s}$ when the measuring junction becomes wet. 

\begin{figure}
  \centering
  \includegraphics[width=0.7\linewidth]{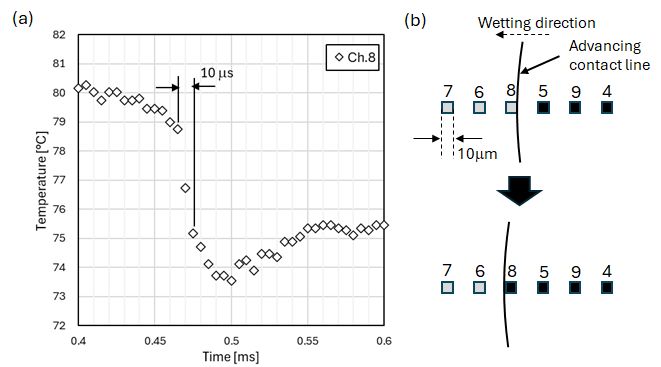}
  \caption{Response to the wetting process}
  \label{Response_wetting}
\end{figure}

The sharp temperature drop coincides with the time when the junction of each thermocouple channel becomes wet. 
The response time of the thermocouple is approximately $10~\mu \mathrm{s}$, corresponding to the duration of the advancing contact line pass through the junction, as shown in Fig.~\ref{Response_wetting}.
Some cases observed a slight temperature increase upon wetting (e.g. Ch.5 in Fig.~\ref{ControlCase}). 
This is not physical since there is no reason for the temperature increase, but it may be due to a non-uniformity of the thermocouple materials. 
A basic principle of thermocouple is that any temperature change in wiring does not influence the voltage and is only applicable when the material characteristics of the wiring is completely uniform.
The inverse voltage may be observed when the drop wets parts of the thermocouple's wiring before the measuring junction.

The contact temperature between the drop and the substrate $T_\mathrm{cont}$ is estimated by assuming two-infinite solid bodies contact each other as $T_\mathrm{cont} = \frac{e_\mathrm{w}T_\mathrm{0}+e_\mathrm{l}T_\mathrm{1}}{e_\mathrm{w}+e_\mathrm{l}}$, where $e_\mathrm{w}$ and $e_\mathrm{l}$ are the thermal effusivity of the substrate and the liquid. 
Note that convection was not included in this calculation. 
The thickness of thermal boundary layers formed in the wafer ($ \sqrt{\alpha_\mathrm{l} t}$) and the water drop ($\sqrt{\alpha_\mathrm{w} t}$) at $t~=~10~\mathrm{ms}$ were estimated to be $320~\mu\mathrm{m}$ and $13~\mu\mathrm{m}$, respectively.
$T_\mathrm{cont}$ was estimated as $71.9\,^\circ\mathrm{C}$ from $T_\mathrm{0}= 81.4\,^\circ\mathrm{C}$ (average of all TC channels) and $T_\mathrm{0}= 12.3\,^\circ\mathrm{C}$ (measured temperature of the needle cooler).
The temperature of Ch.4 and 6-8 (i.e.. except for Ch.~5) indicates a temperature undershoot. 
For Ch.~4, the surface temperature decreases below $T_\mathrm{cont}$. 
The internal liquid flow during drop impact on a hot surface was numerically estimated by \citet{Pasandideh-Fard&Chandra2001}, where downward (out-of-plane) liquid flow occurred around the impact center.
The undershoot was attributed to heat transfer enhancement in the liquid phase by convection. 
The downward flow occurred widely along the wetted region of the spreading drop, so a temperature undershoot could also expected away from the impact center.
The level after convergence was not the same for all channels. 
After the undershoot, the surface temperature increased, and converged. 
The level after converging tended to be higher with the distance of the location from the center.
The levels for Chs.6-8 are higher than $T_\mathrm{cont}$ because of the increase in the liquid temperature.
For those points, the temperature of liquid coming into contact is higher than the initial drop temperature.    

Interestingly, the temperature of Ch.5 had a second temperature drop at $t=~7~\mathrm{ms}$, when the top of the drop moved within the drop body (see the insets of Fig.~\ref{Response_wetting}(c). 
It was triggered by downward, internal liquid flow due to the movement of the top of the drop. 


\section{\label{sec:level4} \emph{Magic carpet breakup} in a depressurized environment}
\subsection{\label{sec:level4.1}Wetted region of the substrate under a spreading drop}
\begin{figure}
  \centering
  \includegraphics[width=1.0\linewidth]{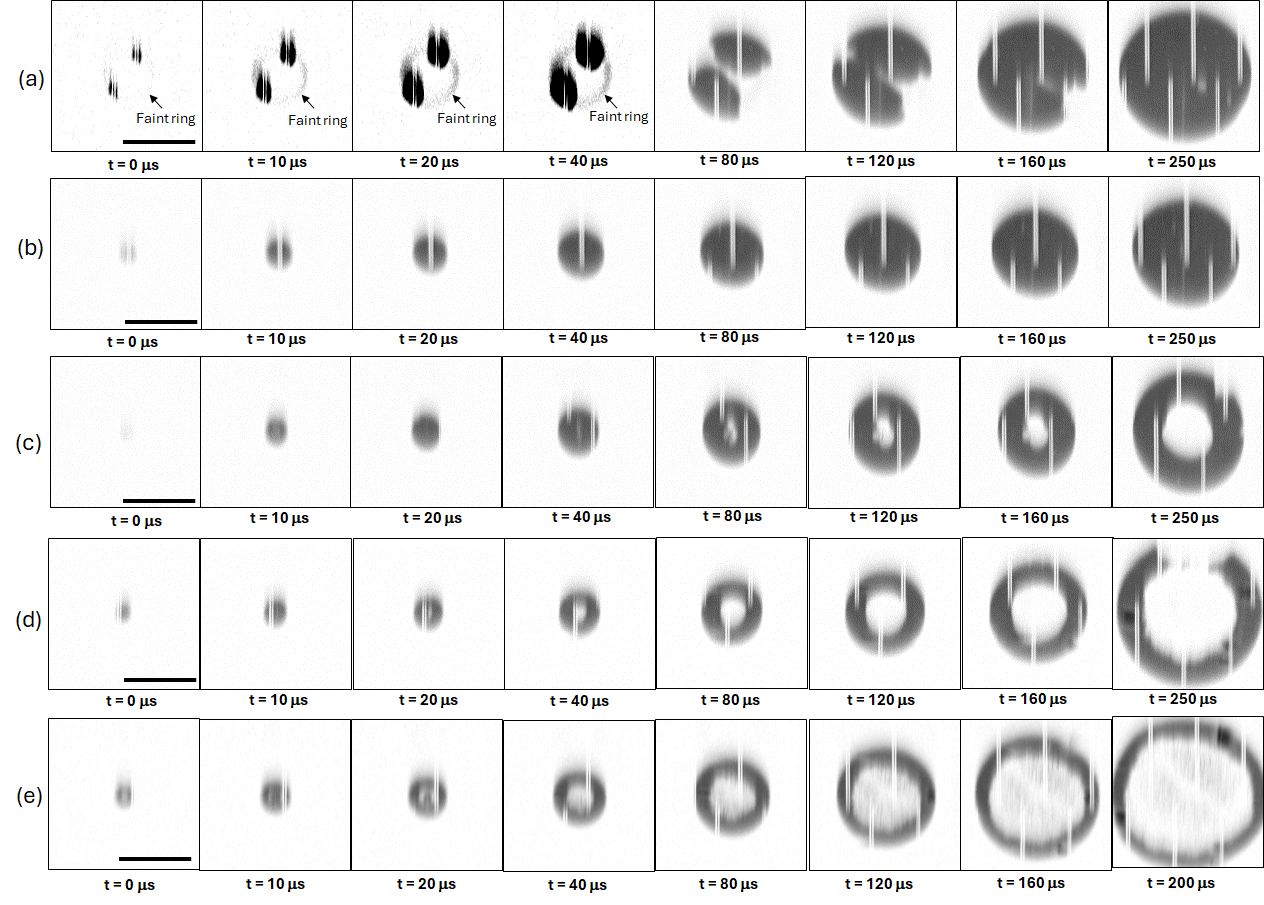}
  \caption{Initial evolution of the wetted region between a water drop and the heated sapphire wafer with the thin-film thermocouple array:(a) $p_\mathrm{ch}=101~\mathrm{kPa}$, $T_\mathrm{0}=80.9\,^\circ\mathrm{C}$ (supplemental video 1a and 1b),; (b) $p_\mathrm{ch}=2.5\mathrm{kPa}$, $T_\mathrm{0}=81.8\,^\circ\mathrm{C}$(supplemental video 2a and 2b);(c) $p_\mathrm{ch}=2.5\mathrm{kPa}$, $T_\mathrm{0}=121.8\,^\circ\mathrm{C}$(supplemental video 3a and 3c);(d) $p_\mathrm{ch}=2.5~\mathrm{kPa}$, $T_\mathrm{0}=140.3\,^\circ\mathrm{C}$ (supplemental video 4a and 4b);(e) $p_\mathrm{ch}=2.5\mathrm{kPa}$, $T_\mathrm{0}=160.8\,^\circ\mathrm{C}$ (supplemental video 5a and 5b);  The scale bar corresponds to 0.9mm. $R_\mathrm{0} = 0.92-1.4~\mathrm{mm}$, $U_\mathrm{0}=0.31-0.36~\mathrm{m/s}$. The image contrast of t=0, 10, 20, and $40~\mu\mathrm{s}$ in (a) is enhanced to show the faint ring more clearly.}
  \label{Initial_wetting}
\end{figure}

\begin{figure}
  \centering
  \includegraphics[width=0.8\linewidth]{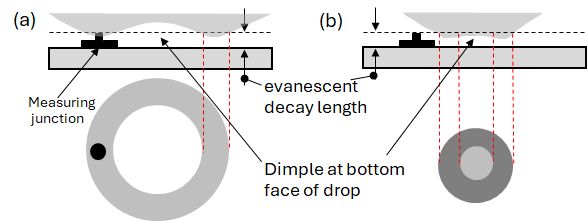}
  \caption{Schematic illustration of the bubble formation, deformation, and growth. The sketch corresponds to the case Fig.~\ref{Initial_wetting}(c-e).}
  \label{illustration_dimple+bump}
\end{figure}

Image sequences of TIR during the initial stages of drop spreading are shown in Fig.~\ref{Initial_wetting} for two different pressures and various wall temperatures. 
The case in Fig.~\ref{Initial_wetting}(a) corresponds to normal atmospheric conditions $p_\mathrm{ch}=101~\rm kPa$, and the cases shown in Fig.~\ref{Initial_wetting}(b-e) are captured under the reduced ambient pressure of $p_\mathrm{ch}=2.5~\mathrm{kPa}$ at various initial substrate temperatures. 

The physical interpretation of the TIR images in Fig.~\ref{Initial_wetting} is illustrated in the sketch in Fig.~\ref{illustration_dimple+bump}. 
The light reflected from the solid-gas (air or vapor) interface decreases exponentially depending on the separation between the impacting fluid and the solid surface with a characteristic decay length of the order of $50~\mathrm{nm}$ \citep{Kolinski2012}.
A faint ring appears first when a drop impacts a smooth surface, indicating a dimple formation at the bottom of the drop.
The ring-shaped region (i.e., closest part) is located within the evanescent length. 
Then, direct contact between fluid and surface occurs at a part of the ring, and then the contact region propagates from it, leading to a spherical wet region with entrainment of an air bubble \citep{Kolinski2012,Lee2012_PRL, Lo&Liu&Xu2017}. 
In the atmosphere, direct contact with the thermocouples occurred almost simultaneously with the appearance of a faint ring, because the thickness of the thermocouples ($45~\mathrm{nm}$ at the measuring junction) is close to the evanescent decay length (typically $50~\mathrm{nm}$ \cite{Kolinski2012}), as shown in Fig.~\ref{illustration_dimple&bump}(a). 

In a depressurized environment ($p_\mathrm{ch}=~2.5~\mathrm{kPa}$), the diameter of the faint ring is smaller than that in the atmosphere. 
Direct contact between the drop and the substrate emerges almost immediately after the appearance of a faint ring, and its outer radius increases monotonically. 
This tendency has been reported by \citep{deGoede&Bonn2019} for isothermal cases in a depressurized environment.
This study showed that the first point to become wet could be on the thermocouple wire or the substrate, depending on the location of the impact center. 
The difference between Fig.~\ref{Initial_wetting}(a) and (b) is due to the difference in the liquid-gas interface profile at the bottom face of the drop; dimple formation is suppressed due to smaller air-cushion effect, resulting in a larger curvature at the bottom (Fig.~\ref{illustration_dimple&bump}(b)). 
This trend persists even when substrate temperature increases further, as shown in Fig.~\ref{Initial_wetting}(c-e). 
The light-gray spot in the wet region (seen in $t=10~\mu s$ of Fig.~\ref{Initial_wetting}(c) and $t=0~\mu s$ of (e)) is a flat-shaped, entrapped gas bubble before shrinkage. 
The initial horizontal radius of the bubble is in the order of $50-100~\mathrm{nm}$, which is consistent with the result in \citep{Li&Hicks&Thoroddsen2017} in a depressurized environment ($p_\mathrm{ch} = 2.8~\mathrm{kPa}$). 
As discussed in \citep{Hatakenaka2019}, based on the threshold temperature between \emph{deposition} and \emph{magic carpet breakup}, the entrapped bubble is considered to act as a nucleate bubble. 
The temperature around the bubble increases to an activation point in a waiting time, and then it grows, as illustrated in Fig.~\ref{Fig1}. 

A single bubble grows considerably from the impact center in a depressurized environment at a relatively high substrate temperature, as shown in Fig.~\ref{Initial_wetting}(c-e). 
A growing bubble at the liquid-solid interface was first observed via TIR by \citet{Yu2019} and by \citet{Hatakenaka2020} afterward.
Both of those works used a smooth substrate. 
The trend of bubble growth and the outward movement of the doughnut-shaped, wet region observed in this study is similar to those observed in \citet{Hatakenaka2020}. 
The wet region continues to move outward even after the ring breaks (i.e., part of the ring becomes dry). 
Partial breakage of the ring always happens somewhere, leading to the drop detachment from the substrate, even when the substrate is smooth.
A small surface defect, drop residue after evaporation, or dust particle determines the location of the breakage.
The presence of a thermocouple may trigger the breakage in this study, but it does not influence the bubble growth behavior or drop outcome.

\subsection{\label{sec:level4.2} Bubble growth at the liquid-solid interface}
\begin{figure}
  \centering
  \includegraphics[width=1.0\linewidth]{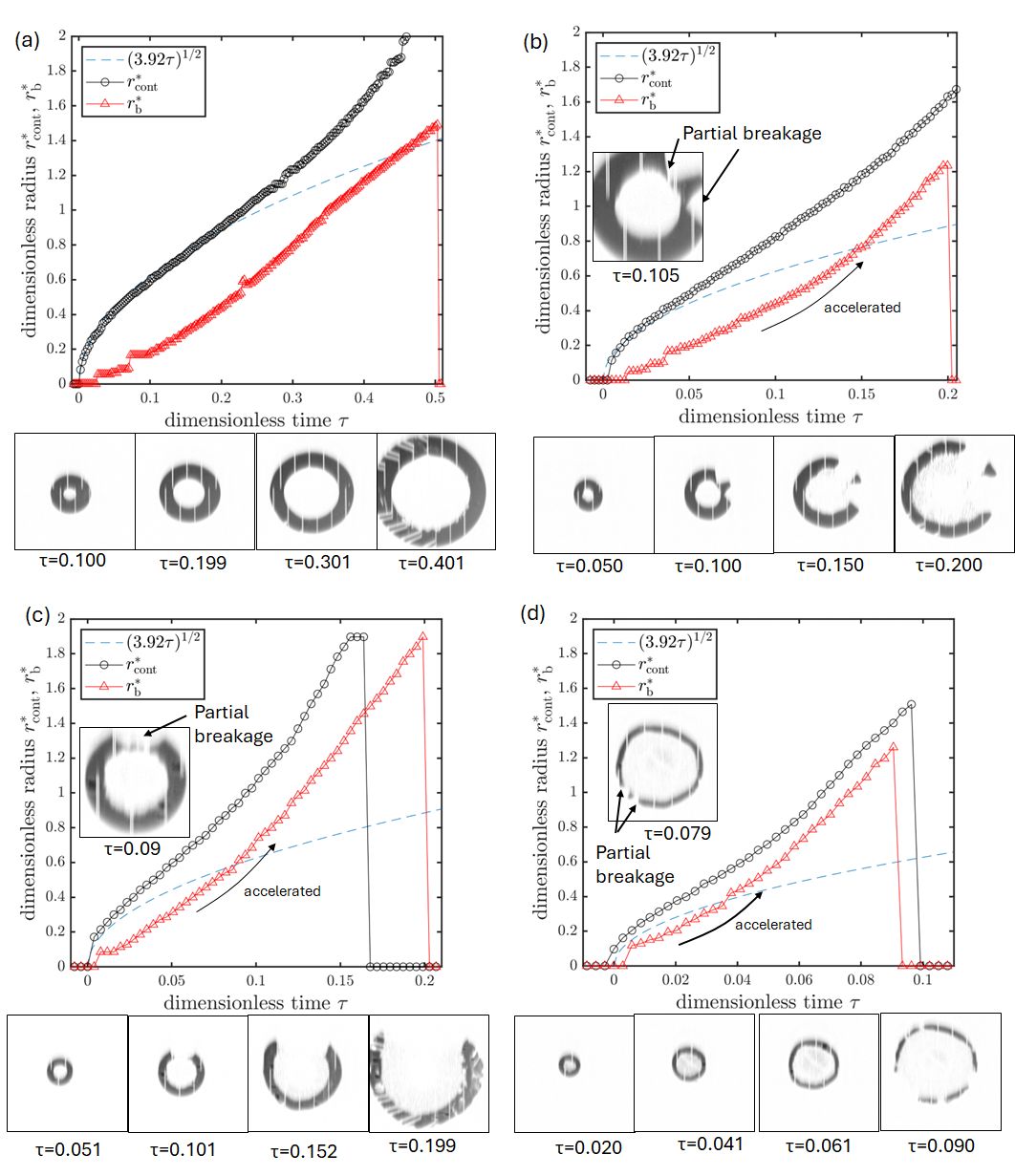}
  \caption{Time evolution of dimensionless bubble radius $R_\mathrm{b} \ast =R_\mathrm{b} / R_\mathrm{0}$ and dimensionless contact radius $R_\mathrm{cont} \ast =R_\mathrm{cont} / R_\mathrm{0}$.  (a) $T_\mathrm{0} = 103\,^\circ\mathrm{C}$,(b) $T_\mathrm{0} = 121.8\,^\circ\mathrm{C}$, (c) $T_\mathrm{0} = 140.3\,^\circ\mathrm{C}$, (d) $T_\mathrm{0} = 160.8\,^\circ\mathrm{C}$. An empirical model for drop spreading in the \emph{deposition} regime \citep{Rioboo&Tropea2002} is shown with the dotted line. The insets and the photos on the bottom are the TIR images subtracted by the background images. The fit circles with the radii of $R_\mathrm{b}$ and $R_\mathrm{cont}$ are overlaid. $p_\mathrm{ch}=2.5~\mathrm{kPa}$.}
  \label{BubbleGrowth+R_cont}
\end{figure}

\begin{figure}
  \centering
  \includegraphics[width=0.7\linewidth]{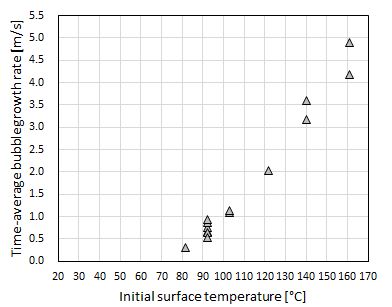}
  \caption{Relationship between time-averaged bubble growth rate and initial surface temperature $T_\mathrm{0}$. $p_\mathrm{ch}=2.5~\mathrm{kPa}$.}
  \label{BubbleGrowthRate}
\end{figure}

The evolution of the dimensionless bubble radius $R^\ast_\mathrm{b} =R_\mathrm{b} / R_\mathrm{0}$ in a depressurized environment ($2.5~\mathrm{kPa}$) for four surface temperatures is shown in Fig.~\ref{BubbleGrowth+R_cont}. 
The horizontal axis is the dimensionless time $\tau = R^\ast_\mathrm{0} / U_\mathrm{0}$, where $R^\ast_\mathrm{0}$ and $U_\mathrm{0}$ are the initial drop radius and the impact velocity evaluated in the side-view images.
The insets are the TIR images subtracted by the background.
Note that $R^\ast_\mathrm{b}$ was also evaluated after a part of the ring-shaped wet region becomes dry.
In Fig.~\ref{BubbleGrowth+R_cont}, the evolution of the dimensionless contact radius $R^\ast_\mathrm{cont} =R_\mathrm{cont} / R_\mathrm{0}$ between the liquid and solid surface, i.e. outer radius of the doughnut-shaped wet region, is compared to an theoretical estimation $R^\ast_\mathrm{cont}= \sqrt{3.92t}$ by \citet{Rioboo&Tropea2002} for a drop spreading on a room temperature surface under atmospheric condition. 
The growing bubble pushed out the liquid, leading to faster outward movement of the contact region. 

The trend of bubble growth is unique; the radius increases almost linearly and accelerates in the later stage rather than asymptotically, a common trend for a bubble growing on a superheated substrate in a liquid pool (i.e., the pool boiling condition) in atmospheric and depressurized conditions.
Interestingly, the bubble growth was not decelerated when partial breakage of the ring-shaped wet region occurred (insets of Fig.~\ref{BubbleGrowth+R_cont}(b-d)).
This indicates that the bubble's internal gas pressure remained higher than the ambient pressure even after the breakage, although the bubble's internal gas could leak to the outside through the opening. 

There are two known modes of bubble growth: inertia-controlled growth and heat-transfer-controlled growth \citep{Carey-Textbook}. 
The Jacob number $\mathrm{Ja}= \frac{c_\mathrm{pl} \, \rho_\mathrm{l} \left( T_\mathrm{0}-T_\mathrm{sat} \right) }{\rho_\mathrm{v} \, L}  \approx \frac{c_\mathrm{pl} \, \rho_\mathrm{l} \, p_\mathrm{0} \, \left(T_\mathrm{0}-T_\mathrm{sat} \right)}{\rho_\mathrm{v0} \, p_\mathrm{ch} \, L}$ of the test conditions for the experimental conditions for \emph{vapor rebound} is extremely high ($Ja=1500-68000$ in \citep{Hatakenaka2019}), so the growth mode is considered to be inertia-dominated throughout the growth process until the drop breaks up.
In this regime, a spherical bubble in an extensive liquid pool grew linearly \citep{Lien1969}. 
\citet{Mikic1970} presented a theoretical model for bubble growth on a superheated surface in a liquid pool, resulting in linear growth in an inertia-dominated regime. 
However, an asymptotic growth curve similar to those in heat-transfer-controlled regime was observed for bubble growth on a superheated surface \citep{vanStralen1975_2, Cole&Shulman1966}. 
The mechanism of asymptotic behavior can be explained by the existence of evaporative liquid microlayer at the bubble base \citep{vanStralen1975_1, Carey-Textbook, Lee&Merte1996_I.J.HMT, Mei&Klausner1995_I.J.HMT1,Mei&Klausner1995_I.J.HMT2}. 
Those experiments were based on bubble growth in a liquid pool, where the surrounding liquid has semi-infinite mass. 
The bubble growth in a drop has rarely been investigated. 
\citet{Hatakenaka2019} presented a theoretical model for bubble growth in an drop impacting a heated surface under \emph{magic carpet breakup} conditions, where limited liquid mass and the influence of drop spreading are considered.
The existence of a liquid microlayer was not considered in the model.
The resulting growth curve was basically linear and slightly accelerated as the bubble grew.
Acceleration occurred because the mass of liquid pushed out by the growing bubble became smaller as the bubble grew.
As discussed in the latter section, a liquid microlayer exists at the bubble base, and its breakup (or hole formation) leads to microdroplet formations. 
The presence of a liquid microlayer should be considered in future growth models.

The relationship between the time-averaged bubble growth rate $\mathrm{d}R_\mathrm{b}/\mathrm{d}t$ and the initial substrate temperature $T_\mathrm{0}$ is shown in Fig.~\ref{BubbleGrowthRate}. 
Note that the saturation temperature ($T_\mathrm{sat}$) of water at the ambient pressure $p_\mathrm{ch}=~2.5~\mathrm{kPa}$ is $21.2\,^\circ\mathrm{C}$. 
Interestingly, the bubble growth rate increases sharply, and its start point is not around the $T_\mathrm{0} = T_\mathrm{sat}$ but at a much higher temperature (approximately $T_\mathrm{0} = 80\,^\circ\mathrm{C}$. 
The driving force of the bubble growth is the difference between vapor pressure in the bubble ($p_\mathrm{v}$) and the ambient pressure ($p_\mathrm{ch}$), whereas the latter is constant in Fig.~\ref{BubbleGrowthRate}. 
The previous model \citep{Hatakenaka2019} predicts that the bubble growth rate has the power law of $R_\mathrm{b}/dt \sim (p_\mathrm{v} - p_\mathrm{ch})^{1/2}$. 
It is reasonable to estimate as  $p_\mathrm{v} = p_\mathrm{sat}(T_\mathrm{v})$ using the saturation curve of the fluid; however, saturation curve's non-linearity does not explain why the bubble growth rate around $T_\mathrm{0} = 80\,^\circ\mathrm{C}$ takes such a small value. 
It is indicated here that the determination of $T_\mathrm{v}$ is a key to understanding this complicated problem. 

\subsection{\label{sec:level4.3} Microdroplets left behind receding contact line}
\begin{figure}
  \centering
  \includegraphics[width=1.0\linewidth]{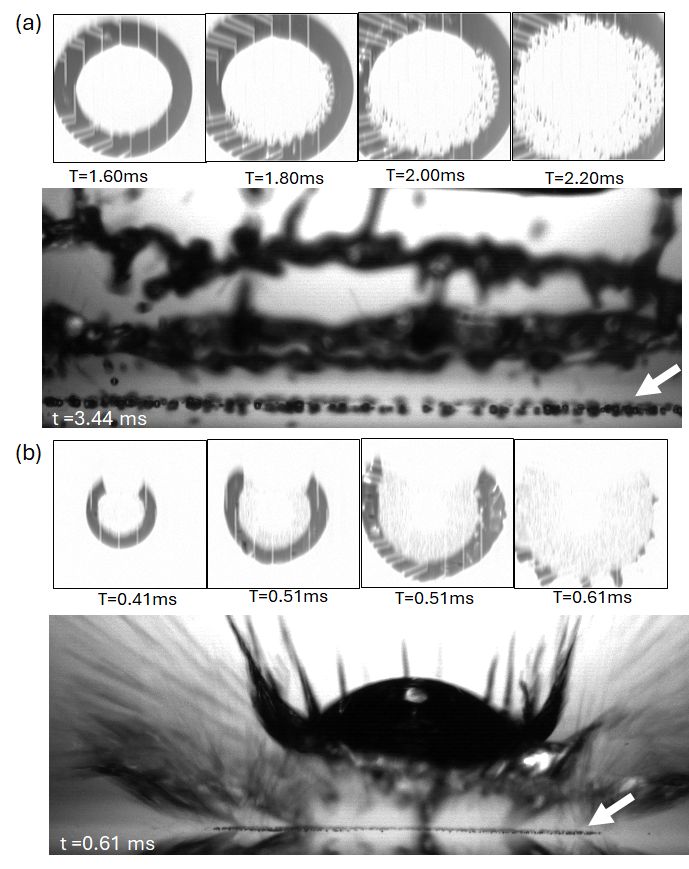}
  \caption{Snapshots and TIR images of liquid microdroplets generated around the growing bubble's receding contact line. $p_\mathrm{ch}~=~2.5~\mathrm{kPa}$. (a)$T_\mathrm{0}~=~103\,^\circ\mathrm{C}$ (supplemental videos 6a and 6b), (b)$T_\mathrm{0}~=~140.3\,^\circ\mathrm{C}$ (supplemental videos 7a and 7b). The TIR images were subtracted by the background images, and image contrast was reduced so that microdroplets became observable. The microdroplets are indicated with white arrows in side-view images.}
  \label{MicroDroplets_pic}
\end{figure}

\begin{figure}
  \centering
  \includegraphics[width=1.0\linewidth]{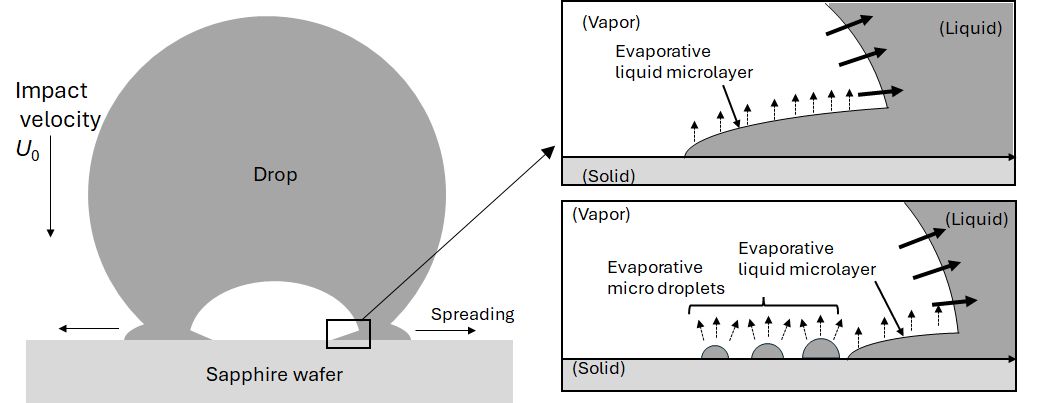}
  \caption{Schematic illustration of evaporative microdroplets formation from evaporative liquid microlyer.}
  \label{MicroDroplets_mechanism}
\end{figure}

The TIR and side-view images around the time of detachment from the surface during \emph{magic carpet breakup} in two initial surface temperatures are shown in Fig.~\ref{MicroDroplets_pic}.
The formation of liquid microdroplets, which were left on the surface after the breakup, was clearly observed in the side-view images.
It was observed in both images that the microdroplets gradually shrink and finally disappear due to evaporation.
The size and location of the microdroplet differed depending on the experimental conditions: large microdroplets were formed in the later stage of bubble growth (just before the breakup) when the surface temperature was low. 
In such cases, the droplets exist around the center of the drop impact. 
On the other hand, for higher surface temperatures, relatively small microdroplets were generated from the beginning of bubble growth.
Thus, the microdroplets distributed all over the bubble base. 
Although determining the microdroplet size is difficult due to limited spatial resolution, the TIR image in Fig.~\ref{MicroDroplets_pic}(b) implies that the droplet size is almost uniform. 

A possible mechanism of microdroplet formation is the breakup of the evaporative liquid microlayer formed at the receding contact line (Fig.~\ref{MicroDroplets_mechanism}). 
The existence of a liquid microlayer beneath a growing bubble has been extensively investigated \citep{Cooper&LLOYD1969, vanStralen1975_1, Utaka2013, Yabuki&Nakabeppu2017}. 
Although there is a so-called contact line model in which evaporation close to the three-phase contact line is dominant, \citet{Fischer&Roisman&Stephan2015} presented that both situations could be observed depending on the experimental parameters.
The liquid microlayer was not directly observed via TIR in this study; however, it is unclear how it appears in the TIR image.
The illumination light would invade the liquid microlayer, reflect at the liquid-vapor interface, and finally reach the image sensor of the TIR camera, if the liquid-vapor interface is parallel to the surface and the liquid thickness is sufficiently small.
That is, the liquid microlayer deposited on a surface may appear in white. 
There are multiple models to estimate the initial thickness profile of the microlayer \citep{Cooper&LLOYD1969, vanStralen1975_1, Utaka2013}. 
A larger bubble growth rate leads to smaller thickness in the model of \citet{Cooper&LLOYD1969, vanStralen1975_1}.
The mechanism of thin liquid film's rapture or hole formation has been studied in \citep{Podgorski2001_PRL, Bonn2009_RevModPhys, Thiele2014_Adv.Col.Int.Sci.}, indicating that the rapture or hole formation tends to happen with smaller disturbance. 
The difference in the location and the microdroplet size observed in Fig.~\ref{MicroDroplets_pic} is attributed to the difference in bubble growth rate. 

Microdroplet generation during \emph{magic carpet breakup} was not always observed in our previous study using smooth substrate \citep{Hatakenaka2019, Hatakenaka2020}; 
Those were only observed in limited test cases. 
Although the difference in imaging condition (the depth of field, the resolution, and noise) may have some influence, the bump of the thermocouple acted as the disturbance. 
This idea needs to be checked using rough surface or patterned surfaces.  


\subsection{\label{sec:level4.4} Surface temperature trend}
\begin{figure}
  \centering
  \includegraphics[width=1.0\linewidth]{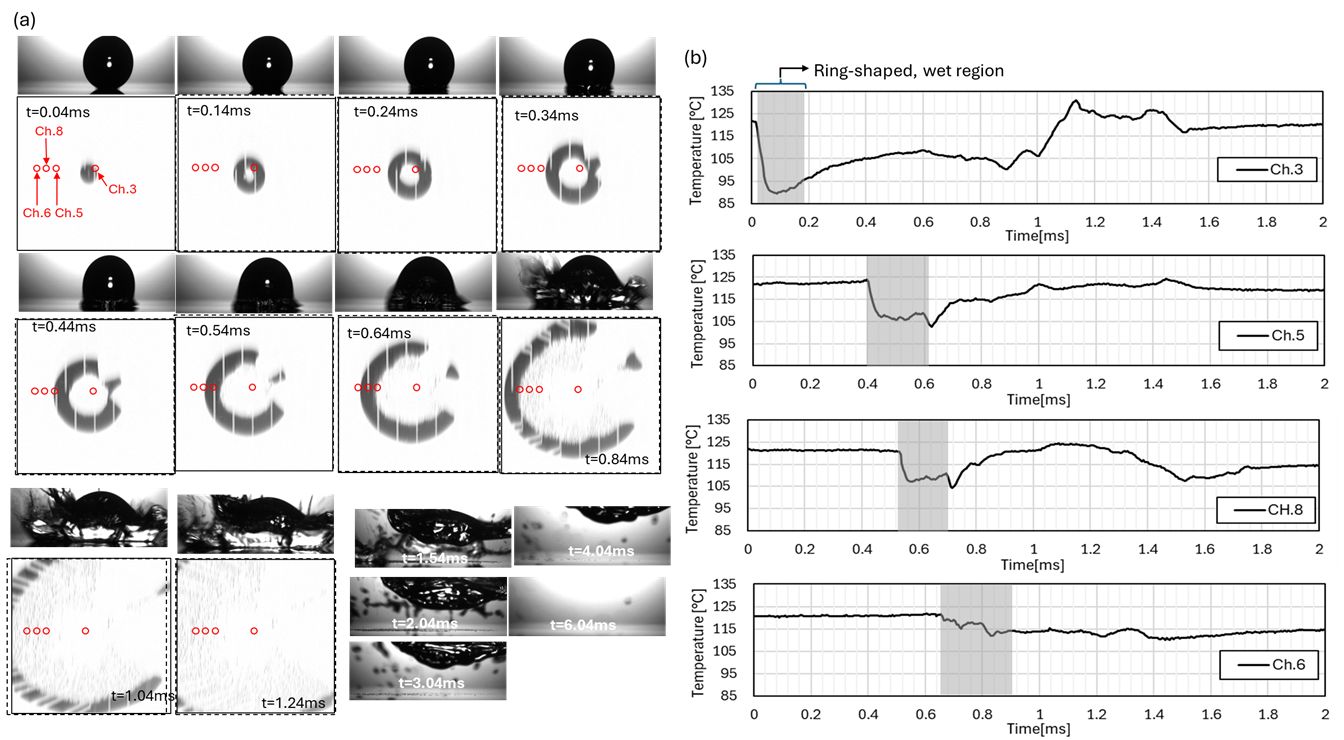}
  \caption{Time evolution of surface temperature compared with the wet region identified via TIR imaging. The TIR images were already subtracted by the background. The time when the ring-shaped wet region covers the location of each TC channel is roughly indicated in the graphs based on TIR images. $p_\mathrm{ch}=2.5~\mathrm{kPa}$, $T_\mathrm{0}=121.8\,^\circ\mathrm{C}$ (supplemental videos 8a and 8b).}
  \label{TempTrend_2500Pa120C+pic}
\end{figure}

\begin{figure}
  \centering
  \includegraphics[width=1.0\linewidth]{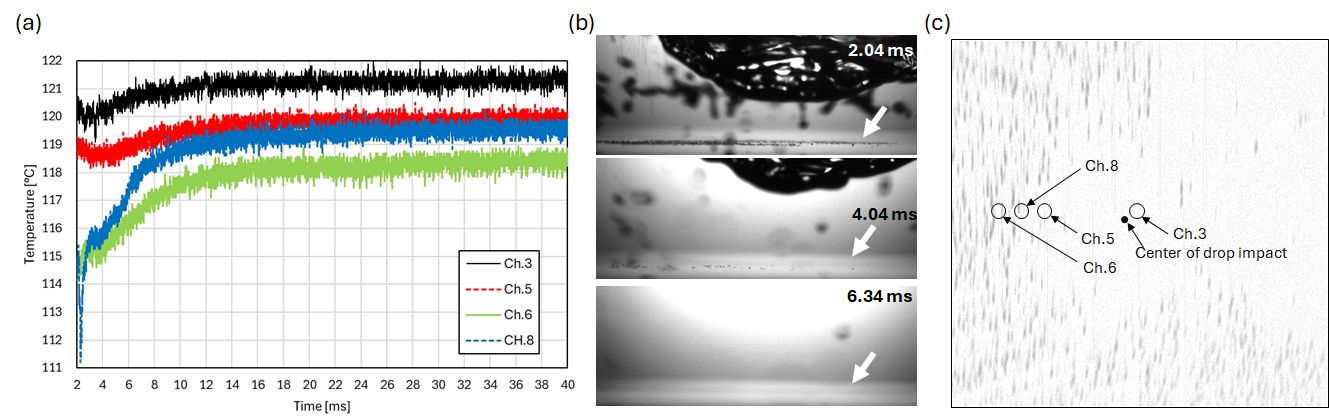}
  \caption{Long-term temperature trend and disappearance of microdroplets on the surface. (a) Long-term temperature trend. (b) Disappearance of microdroplets observed via side-view camera. (c) After background subtraction at $t~=~2.04~\mathrm{ms}$, the TIR image indicates the microdroplet's in-plane distribution. $T_\mathrm{0}=121.8\,^\circ\mathrm{C}$, $p_\mathrm{ch}=2.5~\mathrm{kPa}$ (supplemental videos 8a and 8b).}
  \label{LongTrend+MicroDroplet}
\end{figure}

Exemplary temperature trend during drop impact in a depressurized environment is shown in Figs.~\ref{TempTrend_2500Pa120C+pic} and \ref{LongTrend+MicroDroplet}. Note that the start time in Fig.~\ref{LongTrend+MicroDroplet}(a) corresponds to the end time of Fig.~\ref{TempTrend_2500Pa120C+pic}(b). 

The sharp temperature drops when the measuring junction of each TC becomes wet have been captured similarly to atmospheric conditions.
\citet{Stephan&Hammer1994} proposed a theoretical pool boiling model in which the liquid's detailed thickness profile around the receding contact line is solved by considering an adhesion force, capillary pressure, thermal resistance across the liquid film, and liquid transport at the meniscus. 
Their model predicts a sharp heat flux peak in a narrow region at the meniscus, whose width and thickness are less than $0.2\, \mu \mathrm{m}$ and $0.1\, \mu \mathrm{m}$, respectively.  
\citet{Fischer&Roisman&Stephan2015} reported that the width of the high heat flux region varies depending on the experimental conditions. In the present study, such a sharp temperature drop was not observed around the receding contact line. 
Note that the dimension of the measuring junction is $10$ square micrometer, thus the temperature drop could be significantly dampened even if it exists. 

\begin{figure}
  \centering
  \includegraphics[width=1.0\linewidth]{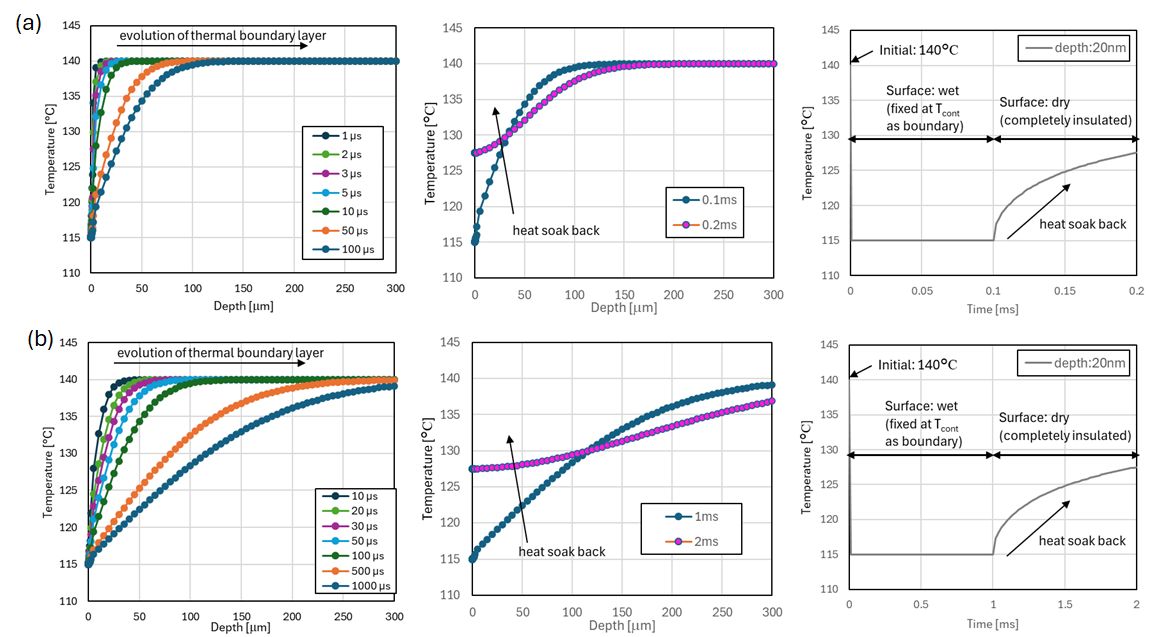}
  \caption{Transient, one-dimensional thermal analysis for a hot substrate when the surface becomes wet, and then turn into dry. The thermal properties of sapphire are used for all solid nodes. The boundary condition at the surface is switched from $T=T_\mathrm{cont}$ to $q~=~\lambda_\mathrm{w}(dT/dz)~=~0$ at a pre-fixed time ($t = ~0.1~\mathrm{ms}$ for (a), and $1~\mathrm{ms}$ for (b))}
  \label{1D-analysis}
\end{figure}

As a reference of typical thermal response when the hot substrate becomes wet and turns into dry, transient, one-dimensional thermal analysis was performed with commercial software (Thermal Desktop/SINDA, C$\&$R Technologies) as shown in Fig.~\ref{1D-analysis}. 
The boundary condition at the surface has been set at $T_\mathrm{0}=T_\mathrm{cont} \sim 115~\,^\circ\mathrm{C}$ for a pre-fixed time ($t = ~0.1~\mathrm{ms}$ for (a), and $1~\mathrm{ms}$, and then switched into $dT/dz = 0$ for simulating the surface becomes dry. 
Here, $z$ is the coordinate in the depth direction of the substrate. 
The thickness of the substrate is $500~\mathrm{nm}$. 
The bottom side of the substrate is fixed at the initial temperature ($140~\,^\circ\mathrm{C}$ as a boundary condition. 
The surface temperature starts to increase due to the heat flow from the deep region of the substrate (heat soak back). 
Despite the difference in the thermal boundary layer's thickness (typically scaled as $\sim \sqrt(t)$), the shape of the curve for temperature recovery due to heat soak back is similar; it starts to increase sharply and then converges to the substrate temperature. 

In Fig.~\ref{TempTrend_2500Pa120C+pic}(b), the time when the location of each thermocouple is wet by the doughnut-shaped, wet region is indicated with a gray band. 
Interestingly, the surface temperature does not increase immediately after the ring-shaped contact region passes through it; rather, it remains at the same temperature level or sometimes decreases further. 
This behavior is attributed to the liquid microdroplets left behind the receding contact line.
As further evidence, in a long-term temperature trend shown in Fig.~\ref{LongTrend+MicroDroplet}(a), the temperature of Channels 6 and 8 remains at a relatively low level at $t=2-6~\mathrm{ms}$, and then starts to increase so that all four channels converges to the initial temperature before drop impact. 
The temperature variation after recovery is due to the error of the calibration test since it remains for a long time until the end of the temperature measurement ($t=80~\mathrm{ms}$).
From the side-view images (Fig.~\ref{TempTrend_2500Pa120C+pic}(b), the microdroplets are confirmed to disappear at around $t=6.34~\mathrm{ms}$. 
The distribution of liquid microdroplets observed via TIR imaging is shown in Fig.~\ref{LongTrend+MicroDroplet}(c).
In this experimental condition, the microdroplet is generated only at the later stage of bubble growth. 
The number of microdroplets around Ch.3 was none or few.
The locations of Channels 5, 6, and 8 are close to each other; however, the number density of droplets is smaller at around Ch.5.
The reason why the surface temperatures around Ch.6 and Ch.8 were relatively lower than the other two (Ch.3 and Ch.5) attributed to the difference in the number of microdroplets around each thermocouple. 

As discussed in \ref{sec:level4.3}, the generation of microdroplets is determined by the bubble growth rate, initial thickness profile, and disturbance triggering the breakup or hole formation. 
The bump of thermocouples acts as the disturbance.
Thus, we need to admit that the phenomenon was unfortunately altered by the existence of thermocouples, even though the height of the bump is only $50~\mathrm{nm}$ at the measurement junction and $20-25~\mathrm{nm}$ for the lead. 
However, the present study's insights can be applied to patterned or rough surface, which is important for industrial application. 
In addition, the existence of a liquid microlayer (or meniscus) and its contribution to the surface temperature evolution after the surface becomes \emph{dry} should also be similar to a smooth substrate.

\section{\label{sec:level5}Conclusion and remarks}
Drop impact onto a superheated substrate in a depressurized environment leads to an explosive drop deformation and backlash motion. 
This study measured a high-speed surface temperature using a thin-film thermocouple array and TIR imaging.
The system successfully captured an extensive bubble growth at the liquid-solid interface and the local temperature change due to the movement of the wet region. 
The surface temperature drops sharply when the location becomes wet; however, no significant temperature drop is detected at the receding contact line. 
This indicates that the heat transfer around the region is not dominant in the heat removal characteristics during this phenomenon. 
On the other hand, microdroplets left behind the receding contact line, and their gradual disappearance due to evaporation have been observed.
The temperature measurement revealed that the surface temperature at the bubble is suppressed even after the receding contact line passes through the location, due to the generation and evaporation of microdroplets.

The only driving force of \emph{vapor rebound} is the vapor pressure in the growing bubble, which is determined by thermodynamic balance at the entire bubble's surface. 
The detailed structure of the triple-phase contact line is especially important. 
The presence of a liquid microlayer at the bubble base is confirmed to be highly probable; however, its detailed thickness profile and the mechanism of liquid microlayer formation are still unknown. 
From a future perspective, the following points are to be discussed. 
First, the existence and thickness distribution of the evaporative liquid microlayer formed at the receding contact line should be directly observed via interferometry imaging technique \citep{Utaka2013}. 
Second, the mechanism of microlayer breakup (or hole formation) and its influence on the heat and mass transfer should be studied and modeled since determining the vapor temperature is key to understand the detailed mechanism of \emph{vapor rebound}. 
Third, the unique bubble growth behavior (i.e., linear bubble growth), which is different from asymptotic growth observed in pool boiling studies, and the dependency of the surface temperature on the bubble growth rate should be studied. 
Understanding these three points sheds light on thermophysical and fluid dynamic modeling, leading to the future acquisition of design and simulation methods for spray cooling technology.
\section*{Acknowledgment}
This work was supported by JSPS KAKENHI Grant Numbers JP20K14676, 20H00223, and 24H00289. A part of this work was supported by "Advanced Research Infrastructure for Materials and Nanotechnology in Japan (ARIM)" of the Ministry of Education, Culture, Sports, Science and Technology (MEXT), Grant Number JPMXP1224AT0149.

\bibliographystyle{elsarticle-num-names}
\bibliography{Reference.bib}

\begin{thebibliography}{85}
\expandafter\ifx\csname natexlab\endcsname\relax\def\natexlab#1{#1}\fi
\providecommand{\url}[1]{\texttt{#1}}
\providecommand{\href}[2]{#2}
\providecommand{\path}[1]{#1}
\providecommand{\DOIprefix}{doi:}
\providecommand{\ArXivprefix}{arXiv:}
\providecommand{\URLprefix}{URL: }
\providecommand{\Pubmedprefix}{pmid:}
\providecommand{\doi}[1]{\href{http://dx.doi.org/#1}{\path{#1}}}
\providecommand{\Pubmed}[1]{\href{pmid:#1}{\path{#1}}}
\providecommand{\bibinfo}[2]{#2}
\ifx\xfnm\relax \def\xfnm[#1]{\unskip,\space#1}\fi
\bibitem[{Kim(2007)}]{Kim2007_Review_spray}
\bibinfo{author}{J.~Kim},
\newblock \bibinfo{title}{Spray cooling heat transfer: The state of the art},
\newblock \bibinfo{journal}{Int. J. Heat Fluid FL} \bibinfo{volume}{28}
  (\bibinfo{year}{2007}) \bibinfo{pages}{753 -- 767}.
  \DOIprefix\doi{https://doi.org/10.1016/j.ijheatfluidflow.2006.09.003}.
\bibitem[{Naber and Reitz(1988)}]{Naber&Reitz1988}
\bibinfo{author}{J.~Naber}, \bibinfo{author}{R.~Reitz},
\newblock \bibinfo{title}{Modeling engine spray/wall impingement},
\newblock \bibinfo{journal}{SAE Transactions} \bibinfo{volume}{97}
  (\bibinfo{year}{1988}) \bibinfo{pages}{118--140}.
\bibitem[{Zhang et~al.(2006)Zhang, Tao, He, and Zhang}]{Zhang2006}
\bibinfo{author}{H.~Zhang}, \bibinfo{author}{W.~Tao}, \bibinfo{author}{Y.~He},
  \bibinfo{author}{W.~Zhang},
\newblock \bibinfo{title}{Numerical study of liquid film cooling in a rocket
  combustion chamber},
\newblock \bibinfo{journal}{International Journal of Heat and Mass Transfer}
  \bibinfo{volume}{49} (\bibinfo{year}{2006}) \bibinfo{pages}{349 -- 358}.
  \URLprefix
  \url{http://www.sciencedirect.com/science/article/pii/S0017931005004369}.
  \DOIprefix\doi{https://doi.org/10.1016/j.ijheatmasstransfer.2005.06.017}.
\bibitem[{Leidenfrost(1966)}]{Leidenfrost1966}
\bibinfo{author}{J.~G. Leidenfrost},
\newblock \bibinfo{title}{On the fixation of water in diverse fire},
\newblock \bibinfo{journal}{International Journal of Heat and Mass Transfer}
  \bibinfo{volume}{9} (\bibinfo{year}{1966}) \bibinfo{pages}{1153 -- 1166}.
  \URLprefix
  \url{http://www.sciencedirect.com/science/article/pii/0017931066901116}.
  \DOIprefix\doi{https://doi.org/10.1016/0017-9310(66)90111-6}.
\bibitem[{Baumeister and Simon(1973)}]{Baumeister1973}
\bibinfo{author}{K.~Baumeister}, \bibinfo{author}{F.~Simon},
\newblock \bibinfo{title}{{L}eidenfrost temperature - its correlation for
  liquid metals, cryogens, hydrocarbons, and water},
\newblock \bibinfo{journal}{Journal of Heat Transfer} \bibinfo{volume}{95}
  (\bibinfo{year}{1973}) \bibinfo{pages}{166--173}.
\bibitem[{Quere(2013)}]{Quere2013}
\bibinfo{author}{D.~Quere},
\newblock \bibinfo{title}{{L}eidenfrost dynamics},
\newblock \bibinfo{journal}{Annu. Rev. Fluid Mech.} \bibinfo{volume}{45}
  (\bibinfo{year}{2013}) \bibinfo{pages}{197--215}.
\bibitem[{Chen and Tseng(1992)}]{Chen1992}
\bibinfo{author}{S.-J. Chen}, \bibinfo{author}{A.~A. Tseng},
\newblock \bibinfo{title}{Spray and jet cooling in steel rolling},
\newblock \bibinfo{journal}{International Journal of Heat and Fluid Flow}
  \bibinfo{volume}{13} (\bibinfo{year}{1992}) \bibinfo{pages}{358 -- 369}.
  \URLprefix
  \url{http://www.sciencedirect.com/science/article/pii/0142727X9290006U}.
  \DOIprefix\doi{https://doi.org/10.1016/0142-727X(92)90006-U}.
\bibitem[{Pola et~al.(2013)Pola, Gelfi, and Vecchia}]{Pola2013}
\bibinfo{author}{A.~Pola}, \bibinfo{author}{M.~Gelfi}, \bibinfo{author}{G.~L.
  Vecchia},
\newblock \bibinfo{title}{Simulation and validation of spray quenching applied
  to heavy forgings},
\newblock \bibinfo{journal}{Journal of Materials Processing Technology}
  \bibinfo{volume}{213} (\bibinfo{year}{2013}) \bibinfo{pages}{2247 -- 2253}.
  \URLprefix
  \url{http://www.sciencedirect.com/science/article/pii/S0924013613002100}.
  \DOIprefix\doi{https://doi.org/10.1016/j.jmatprotec.2013.06.019}.
\bibitem[{Nižetić et~al.(2016)Nižetić, Čoko, Yadav, and
  Grubišić-Čabo}]{Nizetic2016}
\bibinfo{author}{S.~Nižetić}, \bibinfo{author}{D.~Čoko},
  \bibinfo{author}{A.~Yadav}, \bibinfo{author}{F.~Grubišić-Čabo},
\newblock \bibinfo{title}{Water spray cooling technique applied on a
  photovoltaic panel: The performance response},
\newblock \bibinfo{journal}{Energy Conversion and Management}
  \bibinfo{volume}{108} (\bibinfo{year}{2016}) \bibinfo{pages}{287 -- 296}.
  \URLprefix
  \url{http://www.sciencedirect.com/science/article/pii/S0196890415010018}.
  \DOIprefix\doi{https://doi.org/10.1016/j.enconman.2015.10.079}.
\bibitem[{Sargunanathan et~al.(2016)Sargunanathan, Elango, and
  Mohideen}]{Sargunanathan2016}
\bibinfo{author}{S.~Sargunanathan}, \bibinfo{author}{A.~Elango},
  \bibinfo{author}{S.~T. Mohideen},
\newblock \bibinfo{title}{Performance enhancement of solar photovoltaic cells
  using effective cooling methods: A review},
\newblock \bibinfo{journal}{Renewable and Sustainable Energy Reviews}
  \bibinfo{volume}{64} (\bibinfo{year}{2016}) \bibinfo{pages}{382 -- 393}.
  \URLprefix
  \url{http://www.sciencedirect.com/science/article/pii/S1364032116302453}.
  \DOIprefix\doi{https://doi.org/10.1016/j.rser.2016.06.024}.
\bibitem[{Liang and
  Mudawar(2017{\natexlab{a}})}]{Liang&Mudawar2017_Review_spray1}
\bibinfo{author}{G.~Liang}, \bibinfo{author}{I.~Mudawar},
\newblock \bibinfo{title}{Review of spray cooling – part 1: Single-phase and
  nucleate boiling regimes, and critical heat flux},
\newblock \bibinfo{journal}{International Journal of Heat and Mass Transfer}
  \bibinfo{volume}{115} (\bibinfo{year}{2017}{\natexlab{a}})
  \bibinfo{pages}{1174 -- 1205}.
\bibitem[{Liang and
  Mudawar(2017{\natexlab{b}})}]{Liang&Mudawar2017_Review_spray2}
\bibinfo{author}{G.~Liang}, \bibinfo{author}{I.~Mudawar},
\newblock \bibinfo{title}{Review of spray cooling – part 2: High temperature
  boiling regimes and quenching applications},
\newblock \bibinfo{journal}{International Journal of Heat and Mass Transfer}
  \bibinfo{volume}{115} (\bibinfo{year}{2017}{\natexlab{b}})
  \bibinfo{pages}{1206 -- 1222}.
\bibitem[{Breitenbach et~al.(2018)Breitenbach, Roisman, and
  Tropea}]{Breitenbach&Roisman&Tropea2018}
\bibinfo{author}{J.~Breitenbach}, \bibinfo{author}{I.~V. Roisman},
  \bibinfo{author}{C.~Tropea},
\newblock \bibinfo{title}{From drop impact physics to spray cooling models: a
  critical review},
\newblock \bibinfo{journal}{Experiments in Fluids} \bibinfo{volume}{59}
  (\bibinfo{year}{2018}) \bibinfo{pages}{55}.
\bibitem[{Liang and Mudawar(2017)}]{Liang&Mudawar2017_Review_drop}
\bibinfo{author}{G.~Liang}, \bibinfo{author}{I.~Mudawar},
\newblock \bibinfo{title}{Review of drop impact on heated walls},
\newblock \bibinfo{journal}{International Journal of Heat and Mass Transfer}
  \bibinfo{volume}{106} (\bibinfo{year}{2017}) \bibinfo{pages}{103 -- 126}.
\bibitem[{Yarin et~al.(2017)Yarin, Roisman, and
  Tropea}]{Yarin_Roisman_Tropea_book_2017}
\bibinfo{author}{A.~L. Yarin}, \bibinfo{author}{I.~V. Roisman},
  \bibinfo{author}{C.~Tropea}, \bibinfo{title}{Collision Phenomena in Liquids
  and Solids}, \bibinfo{publisher}{Cambridge University Press},
  \bibinfo{year}{2017}.
\bibitem[{Bernardin and Mudawar(2002)}]{Bernardin2002}
\bibinfo{author}{J.~Bernardin}, \bibinfo{author}{I.~Mudawar},
\newblock \bibinfo{title}{A cavity activation and bubble growth model of the
  {L}eidenfrost point},
\newblock \bibinfo{journal}{Journal of Heat Transfer} \bibinfo{volume}{124}
  (\bibinfo{year}{2002}) \bibinfo{pages}{864--874}.
\bibitem[{Bernardin and Mudawar(2004)}]{Bernardin2004}
\bibinfo{author}{J.~Bernardin}, \bibinfo{author}{I.~Mudawar},
\newblock \bibinfo{title}{A {L}eidenfrost point model for impinging droplets
  and sprays},
\newblock \bibinfo{journal}{Journal of Heat Transfer} \bibinfo{volume}{126}
  (\bibinfo{year}{2004}) \bibinfo{pages}{272--278}.
  \DOIprefix\doi{10.1115/1.1652045}.
\bibitem[{Aursand et~al.(2018)Aursand, Davis, and Ytrehus}]{Aursand2018}
\bibinfo{author}{E.~Aursand}, \bibinfo{author}{S.~H. Davis},
  \bibinfo{author}{T.~Ytrehus},
\newblock \bibinfo{title}{Thermocapillary instability as a mechanism for film
  boiling collapse},
\newblock \bibinfo{journal}{Journal of Fluid Mechanics} \bibinfo{volume}{852}
  (\bibinfo{year}{2018}) \bibinfo{pages}{283–312}.
  \DOIprefix\doi{10.1017/jfm.2018.545}.
\bibitem[{Lee et~al.(2020)Lee, Harth, Rump, Kim, Lohse, Fezzaa, and
  Je}]{LEE&Harth&Lohse2020_SoftMatter}
\bibinfo{author}{S.~H. Lee}, \bibinfo{author}{K.~Harth},
  \bibinfo{author}{M.~Rump}, \bibinfo{author}{M.~Kim},
  \bibinfo{author}{D.~Lohse}, \bibinfo{author}{K.~Fezzaa},
  \bibinfo{author}{J.~H. Je},
\newblock \bibinfo{title}{Drop impact on hot plates: contact times{,} lift-off
  and the lamella rupture},
\newblock \bibinfo{journal}{Soft Matter} \bibinfo{volume}{16}
  (\bibinfo{year}{2020}) \bibinfo{pages}{7935--7949}. \URLprefix
  \url{http://dx.doi.org/10.1039/D0SM00459F}.
  \DOIprefix\doi{10.1039/D0SM00459F}.
\bibitem[{Xu et~al.(2005)Xu, Zhang, and Nagel}]{Xu2005}
\bibinfo{author}{L.~Xu}, \bibinfo{author}{W.~W. Zhang}, \bibinfo{author}{S.~R.
  Nagel},
\newblock \bibinfo{title}{Drop splashing on a dry smooth surface},
\newblock \bibinfo{journal}{Phys. Rev. Lett.} \bibinfo{volume}{94}
  (\bibinfo{year}{2005}) \bibinfo{pages}{184505}.
\bibitem[{Driscoll et~al.(2010)Driscoll, Stevens, and
  Nagel}]{Driscol_Stevens_Nagel2010}
\bibinfo{author}{M.~M. Driscoll}, \bibinfo{author}{C.~S. Stevens},
  \bibinfo{author}{S.~R. Nagel},
\newblock \bibinfo{title}{Thin film formation during splashing of viscous
  liquids},
\newblock \bibinfo{journal}{Physical Review E} \bibinfo{volume}{82}
  (\bibinfo{year}{2010}) \bibinfo{pages}{036302}. \URLprefix
  \url{https://link.aps.org/doi/10.1103/PhysRevE.82.036302}.
  \DOIprefix\doi{10.1103/PhysRevE.82.036302}.
\bibitem[{Driscoll and Nagel(2011)}]{Driscoll2011}
\bibinfo{author}{M.~Driscoll}, \bibinfo{author}{S.~Nagel},
\newblock \bibinfo{title}{Ultrafast interference imaging of air in splashing
  dynamics},
\newblock \bibinfo{journal}{Physical Review Letters} \bibinfo{volume}{107}
  (\bibinfo{year}{2011}) \bibinfo{pages}{154502}.
\bibitem[{de~Ruiter et~al.(2012)de~Ruiter, Oh, van~den Ende, and
  Mugele}]{deRuiter2012}
\bibinfo{author}{J.~de~Ruiter}, \bibinfo{author}{J.~M. Oh},
  \bibinfo{author}{D.~van~den Ende}, \bibinfo{author}{F.~Mugele},
\newblock \bibinfo{title}{Dynamics of collapse of air films in drop impact},
\newblock \bibinfo{journal}{Physical Review Letters} \bibinfo{volume}{108}
  (\bibinfo{year}{2012}) \bibinfo{pages}{074505}.
\bibitem[{de~Ruiter et~al.(2015{\natexlab{a}})de~Ruiter, Mugele, and van~den
  Ende}]{deRuiter2015_1}
\bibinfo{author}{J.~de~Ruiter}, \bibinfo{author}{F.~Mugele},
  \bibinfo{author}{D.~van~den Ende},
\newblock \bibinfo{title}{Air cushioning in droplet impact. i. dynamics of thin
  films studied by dual wavelength reflection interference microscopy},
\newblock \bibinfo{journal}{Physics of Fluids} \bibinfo{volume}{27}
  (\bibinfo{year}{2015}{\natexlab{a}}) \bibinfo{pages}{012104}.
\bibitem[{de~Ruiter et~al.(2015{\natexlab{b}})de~Ruiter, van~den Ende, and
  Mugele}]{deRuiter2015_2}
\bibinfo{author}{J.~de~Ruiter}, \bibinfo{author}{D.~van~den Ende},
  \bibinfo{author}{F.~Mugele},
\newblock \bibinfo{title}{Air cushioning in droplet impact. ii. experimental
  characterization of the air film evolution},
\newblock \bibinfo{journal}{Physics of Fluids} \bibinfo{volume}{27}
  (\bibinfo{year}{2015}{\natexlab{b}}) \bibinfo{pages}{012105}.
\bibitem[{de~Ruiter et~al.(2015{\natexlab{c}})de~Ruiter, Lagraaw, van~den Ende,
  and Mugele}]{deRuiter2015_3}
\bibinfo{author}{J.~de~Ruiter}, \bibinfo{author}{R.~Lagraaw},
  \bibinfo{author}{D.~van~den Ende}, \bibinfo{author}{F.~Mugele},
\newblock \bibinfo{title}{Wettability-independent bouncing on flat surfaces
  mediated by thin air films},
\newblock \bibinfo{journal}{Nature Physics} \bibinfo{volume}{11}
  (\bibinfo{year}{2015}{\natexlab{c}}).
\bibitem[{Li and Thoroddsen(2015)}]{Li&Thoroddsen2015}
\bibinfo{author}{E.~Q. Li}, \bibinfo{author}{S.~T. Thoroddsen},
\newblock \bibinfo{title}{Time-resolved imaging of a compressible air disc
  under a drop impacting on a solid surface},
\newblock \bibinfo{journal}{Journal of Fluid Mechanics} \bibinfo{volume}{780}
  (\bibinfo{year}{2015}) \bibinfo{pages}{636–648}.
  \DOIprefix\doi{10.1017/jfm.2015.466}.
\bibitem[{Li et~al.(2017)Li, Langley, Tian, Hicks, and
  Thoroddsen}]{Li&Hicks&Thoroddsen2017}
\bibinfo{author}{E.~Q. Li}, \bibinfo{author}{K.~R. Langley},
  \bibinfo{author}{Y.~S. Tian}, \bibinfo{author}{P.~D. Hicks},
  \bibinfo{author}{S.~T. Thoroddsen},
\newblock \bibinfo{title}{Double contact during drop impact on a solid under
  reduced air pressure},
\newblock \bibinfo{journal}{Phys. Rev. Lett.} \bibinfo{volume}{119}
  (\bibinfo{year}{2017}) \bibinfo{pages}{214502}. \URLprefix
  \url{https://link.aps.org/doi/10.1103/PhysRevLett.119.214502}.
  \DOIprefix\doi{10.1103/PhysRevLett.119.214502}.
\bibitem[{Mandre et~al.(2009)Mandre, Mani, and Brenner}]{Mandre2009}
\bibinfo{author}{S.~Mandre}, \bibinfo{author}{M.~Mani}, \bibinfo{author}{M.~P.
  Brenner},
\newblock \bibinfo{title}{Precursors to splashing of liquid droplets on a solid
  surface},
\newblock \bibinfo{journal}{Physical Review Letters} \bibinfo{volume}{102}
  (\bibinfo{year}{2009}) \bibinfo{pages}{134502}.
\bibitem[{Mani et~al.(2010)Mani, Mandre, and Brenner}]{Mani2010}
\bibinfo{author}{M.~Mani}, \bibinfo{author}{S.~Mandre},
  \bibinfo{author}{M.~Brenner},
\newblock \bibinfo{title}{Events before droplet splashing on a solid surface},
\newblock \bibinfo{journal}{J. Fluid Mech.} \bibinfo{volume}{647}
  (\bibinfo{year}{2010}) \bibinfo{pages}{163--185}.
\bibitem[{Buchm{\"u}ller(2014)}]{Buchmuller2014}
\bibinfo{author}{I.~Buchm{\"u}ller}, \bibinfo{title}{Influence of pressure on
  {L}eidenfrost effect}, Ph.D. thesis, Technische Universit{\"a}t Darmstadt,
  \bibinfo{year}{2014}.
\bibitem[{Celestini et~al.(2013)Celestini, Frisch, and Pomeau}]{Celestini2013}
\bibinfo{author}{F.~Celestini}, \bibinfo{author}{T.~Frisch},
  \bibinfo{author}{Y.~Pomeau}, \bibinfo{title}{Room temperature water
  leidenfrost droplets}, \bibinfo{year}{2013}.
  \href{http://arxiv.org/abs/1308.0941}{{\tt arXiv:1308.0941}}.
\bibitem[{Orejon et~al.(2014)Orejon, Sefiane, and Takata}]{Orejon2014}
\bibinfo{author}{D.~Orejon}, \bibinfo{author}{K.~Sefiane},
  \bibinfo{author}{Y.~Takata},
\newblock \bibinfo{title}{Effect of ambient pressure on {L}eidenfrost
  temprerature},
\newblock \bibinfo{journal}{Phys. Rev. E} \bibinfo{volume}{90}
  (\bibinfo{year}{2014}) \bibinfo{pages}{053012}.
\bibitem[{van Limbeek et~al.(2018)van Limbeek, Hoefnagels, Shirota, Sun, and
  Lohse}]{vanLimbeek2018}
\bibinfo{author}{M.~A.~J. van Limbeek}, \bibinfo{author}{P.~B.~J. Hoefnagels},
  \bibinfo{author}{M.~Shirota}, \bibinfo{author}{C.~Sun},
  \bibinfo{author}{D.~Lohse},
\newblock \bibinfo{title}{Boiling regimes of impacting drops on a heated
  substrate under reduced pressure},
\newblock \bibinfo{journal}{Phys. Rev. Fluids} \bibinfo{volume}{3}
  (\bibinfo{year}{2018}) \bibinfo{pages}{053601}. \URLprefix
  \url{https://link.aps.org/doi/10.1103/PhysRevFluids.3.053601}.
  \DOIprefix\doi{10.1103/PhysRevFluids.3.053601}.
\bibitem[{Breitenbach et~al.(2017)Breitenbach, Roisman, and
  Tropea}]{Breitenbach2017PRL}
\bibinfo{author}{J.~Breitenbach}, \bibinfo{author}{I.~V. Roisman},
  \bibinfo{author}{C.~Tropea},
\newblock \bibinfo{title}{Drop collision with a hot, dry solid substrate: Heat
  transfer during nucleate boiling},
\newblock \bibinfo{journal}{Phys. Rev. Fluids} \bibinfo{volume}{2}
  (\bibinfo{year}{2017}) \bibinfo{pages}{074301}.
\bibitem[{van Limbeek et~al.(2017)van Limbeek, Hoefnagels, Sun, and
  Lohse}]{vanLimbeek2017Soft}
\bibinfo{author}{M.~A.~J. van Limbeek}, \bibinfo{author}{P.~B.~J. Hoefnagels},
  \bibinfo{author}{C.~Sun}, \bibinfo{author}{D.~Lohse},
\newblock \bibinfo{title}{Origin of spray formation during impact on heated
  surfaces},
\newblock \bibinfo{journal}{Soft Matter} \bibinfo{volume}{13}
  (\bibinfo{year}{2017}) \bibinfo{pages}{7514 -- 7520}.
\bibitem[{Hatakenaka et~al.(2019)Hatakenaka, Breitenbach, Roisman, Tropea, and
  Tagawa}]{Hatakenaka2019}
\bibinfo{author}{R.~Hatakenaka}, \bibinfo{author}{J.~Breitenbach},
  \bibinfo{author}{I.~V. Roisman}, \bibinfo{author}{C.~Tropea},
  \bibinfo{author}{Y.~Tagawa},
\newblock \bibinfo{title}{Magic carpet breakup of a drop impacting onto a
  heated surface in a depressurized environment},
\newblock \bibinfo{journal}{Int. J. Heat Mass Transf.} \bibinfo{volume}{145}
  (\bibinfo{year}{2019}) \bibinfo{pages}{118729}. \URLprefix
  \url{http://www.sciencedirect.com/science/article/pii/S0017931019321891}.
  \DOIprefix\doi{https://doi.org/10.1016/j.ijheatmasstransfer.2019.118729}.
\bibitem[{Hatakenaka et~al.(2018)Hatakenaka, Breitenbach, Roisman, Tropea, and
  Tagawa}]{Hatakenaka2018}
\bibinfo{author}{R.~Hatakenaka}, \bibinfo{author}{J.~Breitenbach},
  \bibinfo{author}{I.~V. Roisman}, \bibinfo{author}{C.~Tropea},
  \bibinfo{author}{Y.~Tagawa},
\newblock \bibinfo{title}{Magic carpet breakup of a drop impacting onto a
  heated surface under reduced pressure},
\newblock \bibinfo{journal}{Proc. 14th International Conference on Liquid
  Atomization and Spray Systems}  (\bibinfo{year}{2018}).
\bibitem[{Yu et~al.(2019)Yu, Hu, Zhang, Xiea, and Luo}]{Yu2019}
\bibinfo{author}{X.~Yu}, \bibinfo{author}{R.~Hu}, \bibinfo{author}{X.~Zhang},
  \bibinfo{author}{B.~Xiea}, \bibinfo{author}{X.~Luo},
\newblock \bibinfo{title}{Explosive bouncing on heated silicon surfaces under
  low ambient pressure},
\newblock \bibinfo{journal}{Soft Matter}  (\bibinfo{year}{2019}).
  \DOIprefix\doi{10.1039/C9SM00455F}.
\bibitem[{Pirat et~al.(2023)Pirat, Cottin-Bizonne, Lee, Ramos, and
  Pierre-Louis}]{Pirat2023_PRF}
\bibinfo{author}{C.~Pirat}, \bibinfo{author}{C.~Cottin-Bizonne},
  \bibinfo{author}{C.~Lee}, \bibinfo{author}{S.~M.~M. Ramos},
  \bibinfo{author}{O.~Pierre-Louis},
\newblock \bibinfo{title}{Depressurization-induced drop breakup through bubble
  growth},
\newblock \bibinfo{journal}{Phys. Rev. Fluids} \bibinfo{volume}{8}
  (\bibinfo{year}{2023}) \bibinfo{pages}{L091601}. \URLprefix
  \url{https://link.aps.org/doi/10.1103/PhysRevFluids.8.L091601}.
  \DOIprefix\doi{10.1103/PhysRevFluids.8.L091601}.
\bibitem[{Tong et~al.(2017)Tong, Qiu, Jin, Sun, and
  Duan}]{Tong2017_Appl.Phys.Let.}
\bibinfo{author}{W.~Tong}, \bibinfo{author}{L.~Qiu}, \bibinfo{author}{J.~Jin},
  \bibinfo{author}{L.~Sun}, \bibinfo{author}{F.~Duan},
\newblock \bibinfo{title}{Unique lift-off of droplet impact on high temperature
  nanotube surfaces},
\newblock \bibinfo{journal}{Applied Physics Letters} \bibinfo{volume}{111}
  (\bibinfo{year}{2017}) \bibinfo{pages}{091605}.
  \DOIprefix\doi{10.1063/1.4994022}.
\bibitem[{Auliano et~al.(2018)Auliano, Fernandino, Zhang, and
  Dorao}]{Auliano2018}
\bibinfo{author}{M.~Auliano}, \bibinfo{author}{M.~Fernandino},
  \bibinfo{author}{P.~Zhang}, \bibinfo{author}{C.~A. Dorao},
\newblock \bibinfo{title}{Water droplet impacting on overheated random si
  nanowires},
\newblock \bibinfo{journal}{International Journal of Heat and Mass Transfer}
  \bibinfo{volume}{124} (\bibinfo{year}{2018}) \bibinfo{pages}{307 -- 318}.
  \URLprefix
  \url{http://www.sciencedirect.com/science/article/pii/S0017931017354558}.
  \DOIprefix\doi{https://doi.org/10.1016/j.ijheatmasstransfer.2018.03.042}.
\bibitem[{Moreau et~al.(2019)Moreau, Colinet, and Dorbolo}]{Moreau2019}
\bibinfo{author}{F.~Moreau}, \bibinfo{author}{P.~Colinet},
  \bibinfo{author}{S.~Dorbolo},
\newblock \bibinfo{title}{Explosive {L}eidenfrost droplets},
\newblock \bibinfo{journal}{Phys. Rev. Fluids} \bibinfo{volume}{4}
  (\bibinfo{year}{2019}) \bibinfo{pages}{013602}. \URLprefix
  \url{https://link.aps.org/doi/10.1103/PhysRevFluids.4.013602}.
  \DOIprefix\doi{10.1103/PhysRevFluids.4.013602}.
\bibitem[{van Limbeek et~al.(2017)van Limbeek, Schaarsberg, Sobac, Rednikov,
  Sun, Colinet, and Lohse}]{vanLimbeek2017_JFM}
\bibinfo{author}{M.~A.~J. van Limbeek}, \bibinfo{author}{M.~H.~K. Schaarsberg},
  \bibinfo{author}{B.~Sobac}, \bibinfo{author}{A.~Rednikov},
  \bibinfo{author}{C.~Sun}, \bibinfo{author}{P.~Colinet},
  \bibinfo{author}{D.~Lohse},
\newblock \bibinfo{title}{Leidenfrost drops cooling surfaces: theory and
  interferometric measurement},
\newblock \bibinfo{journal}{Journal of Fluid Mechanics} \bibinfo{volume}{827}
  (\bibinfo{year}{2017}) \bibinfo{pages}{614–639}.
  \DOIprefix\doi{10.1017/jfm.2017.425}.
\bibitem[{Herbert et~al.(2013)Herbert, Fischer, Roisman, and
  P.Stephan}]{Herbert&Stephan2013}
\bibinfo{author}{S.~Herbert}, \bibinfo{author}{S.~Fischer},
  \bibinfo{author}{T.~G. Roisman}, \bibinfo{author}{P.Stephan},
\newblock \bibinfo{title}{Local heat transfer and phase change phenomena during
  single drop impingement on a hot surface},
\newblock \bibinfo{journal}{International Journal of Heat and Mass Transfer}
  \bibinfo{volume}{61} (\bibinfo{year}{2013}) \bibinfo{pages}{605--614}.
  \DOIprefix\doi{https://doi.org/10.1016/j.ijheatmasstransfer.2013.01.081}.
\bibitem[{Jung et~al.(2016)Jung, Jeong, and Kim}]{Jung&Jeong&Kim2016}
\bibinfo{author}{J.~Jung}, \bibinfo{author}{S.~Jeong},
  \bibinfo{author}{H.~Kim},
\newblock \bibinfo{title}{Investigation of single-droplet/wall collision heat
  transfer characteristics using infrared thermometry},
\newblock \bibinfo{journal}{International Journal of Heat and Mass Transfer}
  \bibinfo{volume}{92} (\bibinfo{year}{2016}) \bibinfo{pages}{774 -- 783}.
  \URLprefix
  \url{http://www.sciencedirect.com/science/article/pii/S0017931015009801}.
  \DOIprefix\doi{https://doi.org/10.1016/j.ijheatmasstransfer.2015.09.050}.
\bibitem[{Qi and Weisensee(2020)}]{Qi&Weisensee2020}
\bibinfo{author}{W.~Qi}, \bibinfo{author}{P.~B. Weisensee},
\newblock \bibinfo{title}{Dynamic wetting and heat transfer during droplet
  impact on bi-phobic wettability-patterned surfaces},
\newblock \bibinfo{journal}{Physics of Fluids} \bibinfo{volume}{32}
  (\bibinfo{year}{2020}) \bibinfo{pages}{067110}.
  \DOIprefix\doi{10.1063/5.0010877}.
\bibitem[{Schmidt et~al.(2021)Schmidt, Hofmann, Tenzer, Breitenbach, Tropea,
  and Roisman}]{Schmidt&Roisman&Tropea2021}
\bibinfo{author}{J.~B. Schmidt}, \bibinfo{author}{J.~Hofmann},
  \bibinfo{author}{F.~Tenzer}, \bibinfo{author}{J.~Breitenbach},
  \bibinfo{author}{C.~Tropea}, \bibinfo{author}{I.~V. Roisman},
\newblock \bibinfo{title}{Thermosuperrepellency of a hot substrate caused by
  vapour percolation},
\newblock \bibinfo{journal}{Commnications Physics}  (\bibinfo{year}{2021}).
  \DOIprefix\doi{10.1038/s42005-021-00680-7}.
\bibitem[{Chaze et~al.(2017)Chaze, Caballina, Castanet, and
  Lemoine}]{Chaze&Lemoine2017}
\bibinfo{author}{W.~Chaze}, \bibinfo{author}{O.~Caballina},
  \bibinfo{author}{G.~Castanet}, \bibinfo{author}{F.~Lemoine},
\newblock \bibinfo{title}{Spatially and temporally resolved measurements of the
  temperature inside droplets impinging on a hot solid surface},
\newblock \bibinfo{journal}{Experiment in Fluids} \bibinfo{volume}{58}
  (\bibinfo{year}{2017}) \bibinfo{pages}{96}.
  \DOIprefix\doi{10.1007/s00348-017-2375-1}.
\bibitem[{Seki et~al.(1978)Seki, Kawamura, and Sanokawa}]{Seki1978}
\bibinfo{author}{M.~Seki}, \bibinfo{author}{H.~Kawamura},
  \bibinfo{author}{K.~Sanokawa},
\newblock \bibinfo{title}{Transient temperature profile of a hot wall due to an
  impinging liquid droplet},
\newblock \bibinfo{journal}{ASME. J. Heat Transfer} \bibinfo{volume}{100}
  (\bibinfo{year}{1978}) \bibinfo{pages}{167--169}.
  \DOIprefix\doi{10.1115/1.3450494}.
\bibitem[{Pasandideh-Fard et~al.(2001)Pasandideh-Fard, Aziz, Chandra, and
  Mostaghimi}]{Pasandideh-Fard&Chandra2001}
\bibinfo{author}{M.~Pasandideh-Fard}, \bibinfo{author}{S.~Aziz},
  \bibinfo{author}{S.~Chandra}, \bibinfo{author}{J.~Mostaghimi},
\newblock \bibinfo{title}{Cooling effectiveness of a water drop impinging on a
  hot surface},
\newblock \bibinfo{journal}{International Journal of Heat and Fluid Flow}
  \bibinfo{volume}{22} (\bibinfo{year}{2001}) \bibinfo{pages}{201--210}.
  \DOIprefix\doi{https://doi.org/10.1016/S0142-727X(00)00086-2}.
\bibitem[{P. and L.(1986)}]{Testa&Nicotra1986}
\bibinfo{author}{T.~P.}, \bibinfo{author}{N.~L.},
\newblock \bibinfo{title}{Influence of pressure on the leidenfrost temperature
  and on extracted heat fluxes in the transient mode and low pressure},
\newblock \bibinfo{journal}{Journal of Heat Transfer} \bibinfo{volume}{108}
  (\bibinfo{year}{1986}) \bibinfo{pages}{916--921}. \URLprefix
  \url{https://cir.nii.ac.jp/crid/1360574094645431424}.
  \DOIprefix\doi{10.1115/1.3247034}.
\bibitem[{Stephan and Hammer(1994)}]{Stephan&Hammer1994}
\bibinfo{author}{P.~Stephan}, \bibinfo{author}{J.~Hammer},
\newblock \bibinfo{title}{A new model for nucleate boiling heat transfer},
\newblock \bibinfo{journal}{Heat and Mass Transfer} \bibinfo{volume}{30}
  (\bibinfo{year}{1994}) \bibinfo{pages}{119--125}. \URLprefix
  \url{https://doi.org/10.1007/BF00715018}. \DOIprefix\doi{10.1007/BF00715018}.
\bibitem[{Carey(2008)}]{Carey-Textbook}
\bibinfo{author}{V.~P. Carey}, \bibinfo{title}{Liquid-vapor phase change
  phenomena, Second Edition}, \bibinfo{publisher}{CRC Press},
  \bibinfo{year}{2008}.
\bibitem[{Cooper and Lloyd(1969)}]{Cooper&LLOYD1969}
\bibinfo{author}{M.~G. Cooper}, \bibinfo{author}{A.~J.~P. Lloyd},
\newblock \bibinfo{title}{The microlayer in nucleate pool boiling},
\newblock \bibinfo{journal}{International Journal of Heat and Mass Transfer}
  \bibinfo{volume}{12} (\bibinfo{year}{1969}) \bibinfo{pages}{895--913}.
\bibitem[{van Stralen et~al.(1975)van Stralen, Sohal, Cole, and
  Sluyter}]{vanStralen1975_1}
\bibinfo{author}{S.~J.~D. van Stralen}, \bibinfo{author}{M.~S. Sohal},
  \bibinfo{author}{R.~Cole}, \bibinfo{author}{W.~M. Sluyter},
\newblock \bibinfo{title}{Bubble growth rates in pure and binary systems:
  Combined effect of relaxation and evaporation microlayers},
\newblock \bibinfo{journal}{Int. J. Heat Mass Transf.} \bibinfo{volume}{18}
  (\bibinfo{year}{1975}) \bibinfo{pages}{453--467}.
\bibitem[{Nakabeppu and Furukawa(2006)}]{Nakabeppu2006}
\bibinfo{author}{O.~Nakabeppu}, \bibinfo{author}{Y.~Furukawa},
\newblock \bibinfo{title}{On heat transfer mechanism of nucleate boiling with
  mems sensors: 1st report, temperature measurement beneath boiling bubble (in
  japanese)},
\newblock \bibinfo{journal}{Transaction of the Japan Society Mechanical
  Engineers, series B} \bibinfo{volume}{72} (\bibinfo{year}{2006})
  \bibinfo{pages}{2532--2539}. \DOIprefix\doi{10.1299/kikaib.72.2532}.
\bibitem[{Tange et~al.(2009)Tange, Takagi, Takemura, and Shoji}]{Tange2009}
\bibinfo{author}{M.~Tange}, \bibinfo{author}{S.~Takagi},
  \bibinfo{author}{F.~Takemura}, \bibinfo{author}{M.~Shoji},
\newblock \bibinfo{title}{Bubble growth and fission on mems heat transfer
  surfaces under subcooled boiling conditions (in japanese)},
\newblock \bibinfo{journal}{Transaction of the Japan Society Mechanical
  Engineers, series B} \bibinfo{volume}{75} (\bibinfo{year}{2009})
  \bibinfo{pages}{1655--1661}. \DOIprefix\doi{10.1299/kikaib.75.756_1655}.
\bibitem[{Yabuki and Nakabeppu(2014)}]{Yabuki&Nakabeppu2014}
\bibinfo{author}{T.~Yabuki}, \bibinfo{author}{O.~Nakabeppu},
\newblock \bibinfo{title}{Heat transfer mechanisms in isolated bubble boiling
  of water observed with mems sensor},
\newblock \bibinfo{journal}{International Journal of Heat and Mass Transfer}
  \bibinfo{volume}{76} (\bibinfo{year}{2014}) \bibinfo{pages}{286 -- 297}.
  \URLprefix
  \url{http://www.sciencedirect.com/science/article/pii/S0017931014003172}.
  \DOIprefix\doi{https://doi.org/10.1016/j.ijheatmasstransfer.2014.04.012}.
\bibitem[{Salvadori et~al.(2006)Salvadori, Vaz, Teixeira, Cattani, and
  Brown}]{Salvadori2006}
\bibinfo{author}{M.~C. Salvadori}, \bibinfo{author}{A.~R. Vaz},
  \bibinfo{author}{F.~S. Teixeira}, \bibinfo{author}{M.~Cattani},
  \bibinfo{author}{I.~G. Brown},
\newblock \bibinfo{title}{Thermoelectric effect in very thin film pt-au
  thermocouples},
\newblock \bibinfo{journal}{Applied Physics Letters} \bibinfo{volume}{88}
  (\bibinfo{year}{2006}) \bibinfo{pages}{133106}. \URLprefix
  \url{https://doi.org/10.1063/1.2189192}. \DOIprefix\doi{10.1063/1.2189192}.
  \href{http://arxiv.org/abs/https://doi.org/10.1063/1.2189192}{{\tt
  arXiv:https://doi.org/10.1063/1.2189192}}.
\bibitem[{Liu et~al.(2011)Liu, Sun, Chen, and Xu}]{Liu2011_IEEE}
\bibinfo{author}{H.~Liu}, \bibinfo{author}{W.~Sun}, \bibinfo{author}{Q.~Chen},
  \bibinfo{author}{S.~Xu},
\newblock \bibinfo{title}{Thin-film thermocouple array for time-resolved local
  temperature mapping},
\newblock \bibinfo{journal}{IEEE Electron Device Letters} \bibinfo{volume}{32}
  (\bibinfo{year}{2011}) \bibinfo{pages}{1606--1608}.
  \DOIprefix\doi{10.1109/LED.2011.2165522}.
\bibitem[{Varrenti et~al.(2011)Varrenti, Zhou, Klock, Chyung, Long, Memik, and
  Grayson}]{Varrenti2011}
\bibinfo{author}{A.~Varrenti}, \bibinfo{author}{C.~Zhou},
  \bibinfo{author}{A.~G. Klock}, \bibinfo{author}{S.~H. Chyung},
  \bibinfo{author}{J.~Long}, \bibinfo{author}{S.~O. Memik},
  \bibinfo{author}{M.~Grayson},
\newblock \bibinfo{title}{Thermal sensing with lithographically patterned
  bimetallic thin-film thermocouples},
\newblock \bibinfo{journal}{IEEE Electron Device Letters} \bibinfo{volume}{32}
  (\bibinfo{year}{2011}) \bibinfo{pages}{818--820}.
  \DOIprefix\doi{10.1109/LED.2011.2136314}.
\bibitem[{Kockert et~al.(2019)Kockert, Mitdank, Zykov, Kowarik, and
  Fischer}]{Kockert&Fischer2019}
\bibinfo{author}{M.~Kockert}, \bibinfo{author}{R.~Mitdank},
  \bibinfo{author}{A.~Zykov}, \bibinfo{author}{S.~Kowarik},
  \bibinfo{author}{S.~F. Fischer},
\newblock \bibinfo{title}{Absolute seebeck coefficient of thin platinum films},
\newblock \bibinfo{journal}{Journal of Applied Physics} \bibinfo{volume}{126}
  (\bibinfo{year}{2019}) \bibinfo{pages}{105106}.
  \DOIprefix\doi{10.1063/1.5101028}.
\bibitem[{Coutts(1971)}]{Coutts1971_Review_ThinSolidFilms}
\bibinfo{author}{T.~Coutts},
\newblock \bibinfo{title}{Electrical conduction in thin continuous films},
\newblock \bibinfo{journal}{Thin Solid Films} \bibinfo{volume}{7}
  (\bibinfo{year}{1971}) \bibinfo{pages}{77--100}.
  \DOIprefix\doi{https://doi.org/10.1016/0040-6090(71)90028-9}.
\bibitem[{Kolinski et~al.(2012)Kolinski, Rubinstein, Mandre, Brenner, Weitz,
  and Mahadevan}]{Kolinski2012}
\bibinfo{author}{J.~M. Kolinski}, \bibinfo{author}{S.~M. Rubinstein},
  \bibinfo{author}{S.~Mandre}, \bibinfo{author}{M.~P. Brenner},
  \bibinfo{author}{D.~A. Weitz}, \bibinfo{author}{L.~Mahadevan},
\newblock \bibinfo{title}{Skating on a film of air: Drops impacting on a
  surface},
\newblock \bibinfo{journal}{Phys. Rev. Lett.} \bibinfo{volume}{108}
  (\bibinfo{year}{2012}) \bibinfo{pages}{074503}. \URLprefix
  \url{https://link.aps.org/doi/10.1103/PhysRevLett.108.074503}.
  \DOIprefix\doi{10.1103/PhysRevLett.108.074503}.
\bibitem[{Shirota et~al.(2017)Shirota, van Limbeek, Lohse, and
  Sun}]{Shirota2017}
\bibinfo{author}{M.~Shirota}, \bibinfo{author}{M.~A.~J. van Limbeek},
  \bibinfo{author}{D.~Lohse}, \bibinfo{author}{C.~Sun},
\newblock \bibinfo{title}{Measuring thin films using quantitative frustrated
  total internal reflection ({FTIR})},
\newblock \bibinfo{journal}{The European Physical Journal E}
  \bibinfo{volume}{40} (\bibinfo{year}{2017}).
\bibitem[{Mog(2003)}]{Moghaddam2003}
\bibinfo{title}{{Fabrication and Testing of a Novel Microelectromechnical
  Device for the Study of Boiling Bubble Dynamics}}, volume
  \bibinfo{volume}{Microelectromechanical Systems} of
  \textit{\bibinfo{series}{ASME International Mechanical Engineering Congress
  and Exposition}}, \bibinfo{year}{2003}.
  \DOIprefix\doi{10.1115/IMECE2003-42303}.
\bibitem[{Lee et~al.(2012)Lee, Weon, Je, and Fezzaa}]{Lee2012_PRL}
\bibinfo{author}{J.~S. Lee}, \bibinfo{author}{B.~M. Weon},
  \bibinfo{author}{J.~H. Je}, \bibinfo{author}{K.~Fezzaa},
\newblock \bibinfo{title}{How does an air film evolve into a bubble during drop
  impact?},
\newblock \bibinfo{journal}{Phys. Rev. Lett.} \bibinfo{volume}{109}
  (\bibinfo{year}{2012}) \bibinfo{pages}{204501}. \URLprefix
  \url{https://link.aps.org/doi/10.1103/PhysRevLett.109.204501}.
  \DOIprefix\doi{10.1103/PhysRevLett.109.204501}.
\bibitem[{Lo et~al.(2017)Lo, Liu, and Xu}]{Lo&Liu&Xu2017}
\bibinfo{author}{H.~Y. Lo}, \bibinfo{author}{Y.~Liu}, \bibinfo{author}{L.~Xu},
\newblock \bibinfo{title}{Mechanism of contact between a droplet and an
  atomically smooth substrate},
\newblock \bibinfo{journal}{Phys. Rev. X} \bibinfo{volume}{7}
  (\bibinfo{year}{2017}) \bibinfo{pages}{021036}. \URLprefix
  \url{https://link.aps.org/doi/10.1103/PhysRevX.7.021036}.
  \DOIprefix\doi{10.1103/PhysRevX.7.021036}.
\bibitem[{de~Goede et~al.(2019)de~Goede, de~Bruin, Shahidzadeh, and
  Bonn}]{deGoede&Bonn2019}
\bibinfo{author}{T.~C. de~Goede}, \bibinfo{author}{K.~G. de~Bruin},
  \bibinfo{author}{N.~Shahidzadeh}, \bibinfo{author}{D.~Bonn},
\newblock \bibinfo{title}{Predicting the maximum spreading of a liquid drop
  impacting on a solid surface: Effect of surface tension and entrapped air
  layer},
\newblock \bibinfo{journal}{Phys. Rev. Fluids} \bibinfo{volume}{4}
  (\bibinfo{year}{2019}) \bibinfo{pages}{053602}. \URLprefix
  \url{https://link.aps.org/doi/10.1103/PhysRevFluids.4.053602}.
  \DOIprefix\doi{10.1103/PhysRevFluids.4.053602}.
\bibitem[{Hatakenaka et~al.(2020)Hatakenaka, Harth, Roisman, Tropea, Lohse, and
  Tagawa}]{Hatakenaka2020}
\bibinfo{author}{R.~Hatakenaka}, \bibinfo{author}{K.~Harth},
  \bibinfo{author}{I.~V. Roisman}, \bibinfo{author}{C.~Tropea},
  \bibinfo{author}{D.~Lohse}, \bibinfo{author}{Y.~Tagawa},
\newblock \bibinfo{title}{Bubble growth during “magic carpet breakup” of a
  drop on a heated substrate},
\newblock in: \bibinfo{booktitle}{Proceedings of 73rd Annual Meeting of the APS
  Division of Fluid Dynamics}, \bibinfo{year}{2020}.
\bibitem[{Rioboo et~al.(2002)Rioboo, Marengo, and Tropea}]{Rioboo&Tropea2002}
\bibinfo{author}{R.~Rioboo}, \bibinfo{author}{M.~Marengo},
  \bibinfo{author}{C.~Tropea},
\newblock \bibinfo{title}{Time evolution of liquid drop impact onto solid, dry
  surfaces},
\newblock \bibinfo{journal}{Exp. Fluids} \bibinfo{volume}{33}
  (\bibinfo{year}{2002}) \bibinfo{pages}{112--124}.
\bibitem[{Lien(1969)}]{Lien1969}
\bibinfo{author}{Y.~C. Lien}, \bibinfo{title}{Bubble Growth Rates at Reduced
  Pressure}, Ph.D. thesis, Massachusetts Institute of Technology,
  \bibinfo{year}{1969}.
\bibitem[{Mikic et~al.(1970)Mikic, Rohsenow, and Griffith}]{Mikic1970}
\bibinfo{author}{B.~B. Mikic}, \bibinfo{author}{M.~Rohsenow},
  \bibinfo{author}{P.~Griffith},
\newblock \bibinfo{title}{On bubble growth rates},
\newblock \bibinfo{journal}{International Journal of Heat and Mass Transfer}
  \bibinfo{volume}{13} (\bibinfo{year}{1970}) \bibinfo{pages}{657--666}.
\bibitem[{van Stralen et~al.(1975)van Stralen, Cole, Sluyter, and
  Sohal}]{vanStralen1975_2}
\bibinfo{author}{S.~J.~D. van Stralen}, \bibinfo{author}{R.~Cole},
  \bibinfo{author}{W.~M. Sluyter}, \bibinfo{author}{M.~S. Sohal},
\newblock \bibinfo{title}{Bubble growth rates in nucleate boiling of water at
  subatmospheric pressures},
\newblock \bibinfo{journal}{Int. J. Heat Mass Transfer} \bibinfo{volume}{18}
  (\bibinfo{year}{1975}) \bibinfo{pages}{655--669}.
\bibitem[{Cole and Shulman(1966)}]{Cole&Shulman1966}
\bibinfo{author}{R.~Cole}, \bibinfo{author}{H.~L. Shulman},
\newblock \bibinfo{title}{Bubble growth rates at high {J}akob numbers},
\newblock \bibinfo{journal}{International Journal of Heat and Mass Transfer}
  \bibinfo{volume}{9} (\bibinfo{year}{1966}) \bibinfo{pages}{1377--1390}.
\bibitem[{Lee and Merte(1996)}]{Lee&Merte1996_I.J.HMT}
\bibinfo{author}{H.~S. Lee}, \bibinfo{author}{H.~Merte},
\newblock \bibinfo{title}{Spherical vapor bubble growth in uniformly
  superheated liquids},
\newblock \bibinfo{journal}{International Journal of Heat and Mass Transfer}
  \bibinfo{volume}{39} (\bibinfo{year}{1996}) \bibinfo{pages}{2427--2447}.
  \URLprefix
  \url{https://www.sciencedirect.com/science/article/pii/0017931095003428}.
  \DOIprefix\doi{https://doi.org/10.1016/0017-9310(95)00342-8}.
\bibitem[{Mei et~al.(1995{\natexlab{a}})Mei, Chen, and
  Klausner}]{Mei&Klausner1995_I.J.HMT1}
\bibinfo{author}{R.~Mei}, \bibinfo{author}{W.~Chen}, \bibinfo{author}{J.~F.
  Klausner},
\newblock \bibinfo{title}{Vapor bubble growth in heterogeneous boiling—i.
  formulation},
\newblock \bibinfo{journal}{International Journal of Heat and Mass Transfer}
  \bibinfo{volume}{38} (\bibinfo{year}{1995}{\natexlab{a}})
  \bibinfo{pages}{909--919}. \URLprefix
  \url{https://www.sciencedirect.com/science/article/pii/0017931094001952}.
  \DOIprefix\doi{https://doi.org/10.1016/0017-9310(94)00195-2}.
\bibitem[{Mei et~al.(1995{\natexlab{b}})Mei, Chen, and
  Klausner}]{Mei&Klausner1995_I.J.HMT2}
\bibinfo{author}{R.~Mei}, \bibinfo{author}{W.~Chen}, \bibinfo{author}{J.~F.
  Klausner},
\newblock \bibinfo{title}{Vapor bubble growth in heterogeneous boiling—ii.
  growth rate and thermal fields},
\newblock \bibinfo{journal}{International Journal of Heat and Mass Transfer}
  \bibinfo{volume}{38} (\bibinfo{year}{1995}{\natexlab{b}})
  \bibinfo{pages}{921--934}. \URLprefix
  \url{https://www.sciencedirect.com/science/article/pii/0017931094001963}.
  \DOIprefix\doi{https://doi.org/10.1016/0017-9310(94)00196-3}.
\bibitem[{Utaka et~al.(2013)Utaka, Kashiwabara, and Ozaki}]{Utaka2013}
\bibinfo{author}{Y.~Utaka}, \bibinfo{author}{Y.~Kashiwabara},
  \bibinfo{author}{M.~Ozaki},
\newblock \bibinfo{title}{Microlayer structure in nucleate boiling of water and
  ethanol at atmospheric pressure},
\newblock \bibinfo{journal}{International Journal of Heat and Mass Transfer}
  \bibinfo{volume}{57} (\bibinfo{year}{2013}) \bibinfo{pages}{222 -- 230}.
  \URLprefix
  \url{http://www.sciencedirect.com/science/article/pii/S0017931012007971}.
  \DOIprefix\doi{https://doi.org/10.1016/j.ijheatmasstransfer.2012.10.031}.
\bibitem[{Yabuki and Nakabeppu(2017)}]{Yabuki&Nakabeppu2017}
\bibinfo{author}{T.~Yabuki}, \bibinfo{author}{O.~Nakabeppu},
\newblock \bibinfo{title}{Microlayer formation characteristics in pool isolated
  bubble boiling of water},
\newblock \bibinfo{journal}{Heat Mass Transfer} \bibinfo{volume}{53}
  (\bibinfo{year}{2017}) \bibinfo{pages}{1745–1750}.
  \DOIprefix\doi{https://doi.org/10.1007/s00231-016-1936-9}.
\bibitem[{Fischer et~al.(2015)Fischer, Gambaryan-Roisman, and
  Stephan}]{Fischer&Roisman&Stephan2015}
\bibinfo{author}{S.~Fischer}, \bibinfo{author}{T.~Gambaryan-Roisman},
  \bibinfo{author}{P.~Stephan},
\newblock \bibinfo{title}{On the development of a thin evaporating liquid film
  at a receding liquid/vapour-interface},
\newblock \bibinfo{journal}{International Journal of Heat and Mass Transfer}
  \bibinfo{volume}{88} (\bibinfo{year}{2015}) \bibinfo{pages}{346--356}.
  \URLprefix
  \url{https://www.sciencedirect.com/science/article/pii/S0017931015004251}.
  \DOIprefix\doi{https://doi.org/10.1016/j.ijheatmasstransfer.2015.04.055}.
\bibitem[{Podgorski et~al.(2001)Podgorski, Flesselles, and
  Limat}]{Podgorski2001_PRL}
\bibinfo{author}{T.~Podgorski}, \bibinfo{author}{J.-M. Flesselles},
  \bibinfo{author}{L.~Limat},
\newblock \bibinfo{title}{Corners, cusps, and pearls in running drops},
\newblock \bibinfo{journal}{Phys. Rev. Lett.} \bibinfo{volume}{87}
  (\bibinfo{year}{2001}) \bibinfo{pages}{036102}. \URLprefix
  \url{https://link.aps.org/doi/10.1103/PhysRevLett.87.036102}.
  \DOIprefix\doi{10.1103/PhysRevLett.87.036102}.
\bibitem[{Bonn et~al.(2009)Bonn, Eggers, Indekeu, Meunier, and
  Rolley}]{Bonn2009_RevModPhys}
\bibinfo{author}{D.~Bonn}, \bibinfo{author}{J.~Eggers},
  \bibinfo{author}{J.~Indekeu}, \bibinfo{author}{J.~Meunier},
  \bibinfo{author}{E.~Rolley},
\newblock \bibinfo{title}{Wetting and spreading},
\newblock \bibinfo{journal}{Rev. Mod. Phys.} \bibinfo{volume}{81}
  (\bibinfo{year}{2009}) \bibinfo{pages}{739--805}. \URLprefix
  \url{https://link.aps.org/doi/10.1103/RevModPhys.81.739}.
  \DOIprefix\doi{10.1103/RevModPhys.81.739}.
\bibitem[{Thiele(2014)}]{Thiele2014_Adv.Col.Int.Sci.}
\bibinfo{author}{U.~Thiele},
\newblock \bibinfo{title}{Patterned deposition at moving contact lines},
\newblock \bibinfo{journal}{Advances in Colloid and Interface Science}
  \bibinfo{volume}{206} (\bibinfo{year}{2014}) \bibinfo{pages}{399--413}.
  \URLprefix
  \url{https://www.sciencedirect.com/science/article/pii/S0001868613001553}.
  \DOIprefix\doi{https://doi.org/10.1016/j.cis.2013.11.002},
  \bibinfo{note}{manuel G. Velarde}.

\end{thebibliography}

\end{document}